\documentclass[a4paper,12pt]{article}
\pdfoutput=1

\usepackage{amssymb} 
\usepackage{amsmath}
\usepackage{epsfig}
\usepackage{graphicx}

\makeatletter

\@addtoreset{equation}{section}
\makeatother
\def\be{\begin{equation}}
\def\ee{\end{equation}}
\def\bea{\begin{eqnarray}}
\def\eea{\end{eqnarray}}

\def\({\left(}
\def\){\right)}
\def\<{\left<}
\def\>{\right>}

\def\tr{{\mbox{tr}}}
\def\be{\begin{equation}}
\def\ee{\end{equation}}
\def\bea{\begin{eqnarray}}
\def\eea{\end{eqnarray}}
\def\ben{\begin{eqnarray}}
\def\een{\end{eqnarray}}
\def\({\left(}
\def\){\right)}
\def\<{\left<}
\def\>{\right>}
\def\!{\right|}
\def\|{\left|}

\def\[{\left[}
\def\]{\right]}

\def\+{\bar}
\def\mb{\mathbb}
\def\tr{{\mbox{tr}}}

\def\L{{\cal{L}}}
\def\t{\widetilde}
\def\A{{\bf{A}}}
\def\B{{\bf{B}}}
\def\C{{\bf{C}}}

\def\M{{\cal{M}}}

\def\N{{\cal{N}}}

\def\Up{{\Upsilon}}

\def\L{{\cal{L}}}

\def\eps{{\cal{\varepsilon}}}
\def\ep{{\epsilon}}

\def\E{{\cal{E}}}

\def\h{\widehat}

\begin{document}

\setlength{\unitlength}{1mm}

\pagestyle{empty}
\vskip-10pt
\vskip-10pt
\hfill %{\tt hep-th/yymmnnn}
\begin{center}
\vskip 3truecm
{\Large \bf
The (1,0) tensor and hypermultiplets in loop space}
\vskip 2truecm
{%\large \bf
\textsc{Dongsu Bak,$^{ \negthinspace  a,b}$  Andreas Gustavsson,$^{ \negthinspace a}$} }
\vskip0.8cm 
\centerline{\sl  a) Physics Department, University of Seoul, Seoul 02504 \rm KOREA}
 \vskip0.4cm
\centerline{\sl  b) Natural Science Research Institute, University of Seoul, Seoul 02504 \rm KOREA}
\vskip 0.7truecm
\begin{center}
(\tt dsbak@uos.ac.kr,\,\,agbrev@gmail.com)
\end{center}
\end{center}
\vskip 2truecm
{\abstract We show that the (1,0) tensor and hypermultiplet supersymmetry variations can be uplifted to loop space. Upon dimensional reduction we make contact with abelian five-dimensional super Yang-Mills, which has a nonabelian generalization that we subsequently uplift back to loop space where we conjecture a nonabelian generalization of the (1,0) supersymmetry variations and demonstrate their on-shell closure.}

\vfill
\vskip4pt
\eject
\pagestyle{plain}

\section{Introduction}
It has for a long time been thought that there shall be a loop space formulation of the $(2,0)$ superconformal M5 brane theory, see for instance \cite{Schreiber:2005ff,Gustavsson:2005aq,Huang:2010db,Papageorgakis:2011xg}. If we compactify one spatial direction on a circle, then we may consider the loops that wrap around the circle. Let $v^M$ be the Killing vector that generates the circle
\ben
\frac{dC^M}{ds} &=& v^M(C(s))\label{minia}
\een
We will refer to the subspace of loops that satisfy (\ref{minia}) as the mini loop space. If we furthermore perform dimensional reduction along the
circle, then the mini loop space formulation should descend to five-dimensional super Yang-Mills (SYM). If we have a Killing vector field $v^M$ on the six-manifold, then we may write down nonabelian supersymmetry variations in a six-dimensional covariant form that correspond to the nonabelian $(2,0)$ superconformal variations, but where one has to impose as a constraint on all fields, collectively denoted as $\Phi$, that their Lie derivatives vanish along the vector field $v^M$,
\bea
\L_v \Phi &=& 0
\eea
There are two different formulations of 5d SYM that are 6d covariant. One formulation \cite{Lambert:2010wm,Gustavsson:2018rcc} uses a nonabelian selfdual field strength $H_{MNP}$. In addition one introduces a one-form gauge potential $A_M$ with field strength $F_{MN}$. This gauge field is needed in order to write gauge covariant derivatives of matter fields. One then finds that closure of the supersymmetry variations requires the constraint $F_{NP} = v^M H_{MNP}$ and moreover $H_{MNP}$ has to satisfy a modified Bianchi identity and be selfdual. By further analysis, it was found in \cite{Gustavsson:2018rcc} that one may express the supersymmetry variations entirely in terms of $A_M$ and its field strength $F_{MN}$ without introducing $H_{MNP}$. Closure of these supersymmetry variations requires the constraints
\bea
\L_v A_M = 0, \ \ \ \ \ %\cr
A_M v^M = 0
\eea
For the abelian case, one may take $A_N = v^M B_{MN}$ which automatically solves the constraint $A_M v^M = 0$. In the nonabelian case we do not have a nonabelian two-form $B_{MN}$ in either formulation. In addition there are five scalar fields $\phi^A$ and four real fermions $\psi$. By supersymmetry, these also have to be constrained by 
\bea
\L_v \phi^A = 0,\ \ \ \ \ \ \ %
%\cr
\L_v \psi = 0
\eea
If the vector field $v^M$ is spacelike, then the supersymmetry variations are given by \cite{Gustavsson:2018rcc}
\ben
\delta \phi^A &=& i \bar\eps \Gamma^A \psi\cr
\delta \psi &=& \frac{1}{2} \Gamma^{MNP} \eps F_{MN} u_P + \Gamma^M \Gamma^A \eps D_M \phi^A - 4 \Gamma^A \eta \phi^A - \frac{1}{2} \Gamma_M \Gamma^{AB} \eps [\phi^A,\phi^B] v^M\cr
\delta A_M &=& i \bar\eps \Gamma_{MN} \psi v^N\label{vsusy}
\een
where $u_M = v_M/g^2$ and $g = \sqrt{v^M v_M}$ and where the supersymmetry parameter is a conformal Killing spinor satisfying
\ben
\nabla_M \eps &=& \Gamma_M \eta\label{Sp2CKE}
\een
as well as
\bea
\L_v \eps &=& 0
\eea
These constraints imply that we have a five-dimensional super-Yang-Mills theory, which has been formulated in a six-dimensional covariant form.

If we interpret $v^M$ as the tangent vector $v^M(C(s)) = \dot{C}^M(s)$ of a loop $s \mapsto C^M(s)$ then this leads to the loop space formulation in \cite{Papageorgakis:2011xg}. A hope has been that what above appeared as dimensional reduction, would in loop space become reparametrization invariance of loop space fields, where the Lie derivative $\L_v$ would become the change of the loop space field under a reparametrization, essentially by
\bea
\frac{dC^M(s)}{ds} \frac{\delta}{\delta C^M(s)} &=& \frac{d}{ds}
\eea
While this idea sounds nice, we do not expect that this approach can be used to fully realize superconformal symmetry in loop space. 

In a loop space formulation, it gets impossible to disentangle the supersymmetry parameter from the local field when both are being integrated along the loop. This is the key observation that motivated us to seek a different loop space formulation. We will use a cohomological form of the supersymmetry variations where the supersymmetry parameter is absorbed in the fermionic fields. We will reduce the amount of supersymmetry down to $(1,0)$ supersymmetry, where we have one tensor multiplet and an arbitrary number of hypermultiplets. We will lift each of these multiplets to loop space and show that this can be done while keeping the $(1,0)$ superconformal symmetry intact, for both abelian and nonabelian gauge groups.

One may ask about a local field realization of the nonabelian loop space fields. We will not address this question here. Perhaps such a formulation does not exist and, instead, the loop space fields describe tensionless strings (or loops) in which case the M5 brane theory would be a string field theory. For strings that wrap around the circle fiber we can make contact with the local fundamental fields in five-dimensional super-Yang-Mills. But for strings that do not wrap around the circle-fiber we may get a dual formulation of five-dimensional super-Yang-Mills. Keeping both wrapped and unwrapped string will give an over-count of degrees of freedom. We would find monopole strings as fundamental string fields as well as solitonic objects, coming from wrapped and unwrapped strings in six dimensions. This has been an argument against a string field description of the $(2,0)$ theory \cite{Seiberg:1997ax}. On the other hand, this problem of over-counting is closely related to selfduality. Our analysis in this paper is at the classical level. To properly deal with selfduality, we should discuss the quantum theory. One way to do this is by starting with a nonchiral theory and perform holomorphic factorization of the resulting partition function \cite{Witten:1996hc}. 

Our classical description in loop space may be larger than the desired $(1,0)$ theory that we are after. But still, having a classical description can be interesting if it contains some aspects of the $(1,0)$ theory. Having a string field theory description does not necessarily mean that the theory would be nonlocal. We expect the $(1,0)$ theories to have a local stress tensor, but the description of these theories may appear to be nonlocal when it is expressed in terms of loop space fields \cite{Seiberg:1997zk,Beem:2015aoa}.

The structure of the paper is as follows. In section \ref{section2} we present the motivations for our loop space construction. In section \ref{section3} we obtain the abelian supersymmetry variations for the (1,0) theory by making a particular supersymmetry breaking of the (2,0) theory. In section \ref{section4} we show that there exists a unique lightlike vector $U$ that is linearly independent from the conformal Killing vector $V$ that we construct as the Dirac current of the two supersymmetry parameters $\eps_I$ (for $I = 1,2$) in the (1,0) theory. In section \ref{section5} we perform dimensional reduction down to five-dimensional super Yang-Mills where we present the nonabelian supersymmetry variations first in their ordinary form and next in their cohomological form. In section \ref{section6} we study the tensor multiplet and obtain the supersymmetry variations in loop space, first for abelian gauge group, and next for nonabelian gauge group. In section \ref{section7} we study the hypermultiplet and obtain the supersymmetry variations in loop space, first for abelian gauge group, and next for nonabelian gauge group. In section \ref{section8} we discuss our results and present some future directions. 

There is also an appendix. In section \ref{sectionA} we derive a closed formula for $\L_V \eps_I$. In section \ref{dagger} we introduce $\delta$ and $\delta^{\dag}$ in loop space. In section \ref{UpQAI} we obtain the supersymmetry variation of one particular hypermultiplet fermion field that we denote as $\Up_M^{AI}$ and uplift this supersymmetry variation to loop space. In section \ref{RIJ} we show that the R-symmetry generator that appears in the closure relations of the supersymmetry variations, is covariantly constant. In section \ref{LieDer} we introduce Lie derivatives in loop space.

\section{Roadmap}\label{section2}
In this section we explain and motivate what we do in this paper. We explain why we introduce the loop space fields in the way that we do it, and why we believe that it can not be done in a different way. 

The geometry of the Lorentzian six-manifold on which we can put a classical abelian chiral superconformal $(1,0)$ theory is restricted by the existence of two chiral conformal Killing spinors $\eps_I$ for $I = 1,2$ that satisfy
\bea
\nabla_M \eps_I &=& \Gamma_M \eta_I\label{CKEa}
\eea
The index $I$ transforms in the fundamental representation of the $SU(2)$ R-symmetry. We have the invariant antisymmetric tensor $\ep_{IJ}$ and its inverse $\ep^{IJ}$ by which the indices $I, J,.. $ will be rised and lower from the right, $\psi^I = \psi_J \ep^{JI}$ and $\psi_J = \psi^J \ep_{JI}$.

The classification of Lorentzian six-manifolds with at least one complex chiral conformal Killing spinor is known \cite{Baum}. These are the six-manifolds on which we can put a classical $(1,0)$ superconformal theory, because one complex spinor is equivalent to two real conformal Killing spinors $\eps_I$. However, as we will argue below, the criterion for the  Lorentzian six-manifolds on which we can put a $(1,0)$ superconformal quantum theory, could be a somewhat different criterion. 

From (\ref{CKEa}) we can derive several geometric relations. As it turns out, we will need each of the geometric relations that we summarize below (omitting all proofs) in order to be able to uplift the $(1,0)$ tensor multiplet supersymmetry variations to loop space. Our proofs of these geometric relations will be presented later in the paper and in the appendices. 

From (\ref{CKEa}) we can obtain a corresponding equation for $\eta_I$,
\bea
\nabla_M \eta_I &=& \(\frac{R}{80} g_{MN} - \frac{1}{8} R_{MN}\) \Gamma^N \eps_I
\eea
where our conventions for the curvature tensors are $[\nabla_M,\nabla_N] W_P = R_{MNP}{}^Q W_Q$ (for a vector $W_P$) and $R_{MN} = R_{MPN}{}^P$. From the two spinors $\eps_I$ we can construct a conformal Killing vector $V_M$ and an antiselfdual three-form $\Theta_{MNP,IJ}$, which is symmetric in $I,J$, as
\bea
V^M &=& \bar\eps^I \Gamma^M \eps_I\cr
\Theta_{MNP,IJ} &=& \bar\eps_I \Gamma_{MNP} \eps_J
\eea
These have the properties
\bea
\nabla_M V_N + \nabla_N V_M &=& \frac{\Omega}{3} g_{MN}
\eea
where
\bea
\Omega &=& \nabla_M V^M
\eea
and
\bea
V_M V^M &=& 0\cr
\Theta_{MNP,IJ} V^P &=& 0
\eea
Various other properties of $\Theta_{MNP,IJ}$ are derived in Appendix \ref{APPF}.

There is another independent lightlike vector field $U^M$ (see section \ref{section4} for its detailed construction), which commutes with $V^M$,
\bea
\L_V U^N = [V,U]^N = V^M \nabla_M U^N - U^M \nabla_M V^N = 0
\eea
If we define a normalization factor as 
\bea
\N &=& V^M U_M
\eea
then the ratio $\frac{U^M}{\N}$ is uniquely determined by $V^M$. We define
\bea
\Theta_{MN,IJ} &=& \Theta_{MNP,IJ} U^P\cr
R^I{}_J &=& \bar\eps^I \eta_J - \frac{1}{2} \delta^I_J \bar\eps^K \eta_K\label{F2}
\eea
We have the following Lie derivatives,
\ben
\L_V \eps_I &=& \frac{\Omega}{12} \eps_I + 8 \eps_J R^J{}_I\cr
\L_V \Theta_{MN}{}^I{}_J &=& 8 \(\Theta_{MN}{}^I{}_K R^K{}_J - R^I{}_K \Theta_{MN}{}^K{}_J\) + \frac{1}{3} \Omega \Theta_{MN}{}^I{}_J\label{Lie12}
\een
and importantly, the R-symmetry generator is covariantly constant 
\bea
\nabla_M R^I{}_J &=& 0
\eea
This property enables us to bring $R^I{}_J$ outside an integral over a loop, so it is a crucial property, which is necessary for a loop space formulation to exist. 

These relations, which we will derive from the assumption that we have a Lorentzian six-manifold that supports at least two chiral conformal Killing spinors $\eps_I$, are sufficient for uplifting the tensor multiplet to loop space. 

To be able to uplift the hypermultiplet to loop space however, it turns out that we need in addition $U$ to be a conformal Killing vector,
\bea
\nabla_M U_N + \nabla_N U_M &=& \frac{\Omega^{\vee}}{3} g_{MN}
\eea
where we define
\bea
\Omega^{\vee} &=& \nabla_M U^M
\eea
At a more technical level, we also need the relation
\ben
\L_U \eps_I &=& \frac{\Omega^{\vee}}{12} \eps_I\label{eqB}
\een
that we have not been able to prove, although it is easy to see that by assuming (\ref{eqB}) we get $\L_U V_M = \frac{\Omega^{\vee}}{3} V_M$, which is a relation that we can easily prove by assuming that $U$ is a conformal Killing vector that commutes with $V$. From $[\L_U,\L_V] g_{MN} = 0$ one can see that we have the relation
\ben
\L_V \Omega^{\vee} &=& \L_U \Omega\label{ovee1}
\een
By using (\ref{ovee1}) we can see that (\ref{eqB}) is consistent with $[\L_V,\L_U] \eps_I = 0$. From $\L_U V^M = [U,V]^M = 0$, it is also clear that, by assuming that $U$ is a conformal Killing vector, the most general possibility is to have
\bea
\L_U \eps_I &=& \frac{\Omega^{\vee}}{12} \eps_I + \kappa_I
\eea
where 
\bea
\bar\eps^I \Gamma^M \kappa_I &=& 0
\eea
In section \ref{section4} and appendix \ref{ConfEin} we show that the properties for $U$ that it is a conformal Killing vector that commutes with $V$, can be derived if we assume that there exists two nonchiral conformal Killing vectors $\E_I$ on the six-manifold such that $\bar\E^I \E_I$ is nonzero. The existence of two nonchiral conformal Killing spinors $\E_I$ is necessary for abelian gauge group in order to define the superconformal theory in Euclidean signature where we do not have symplectic Majorana-Weyl spinors. It is also needed if one defines the partition function of a chiral $(1,0)$ superconformal theory by the partition function of the nonchiral $(1,1)$ superconformal theory for which there is a classical Lagrangian, by applying holomorphic factorization following \cite{Witten:1996hc}, \cite{Henningson:1999dm}. For curved spacetimes, we need a complex deformation of the metric away from the Lorentzian metric in order to even just define say the Feynman propagator of a free field theory \cite{Visser:2017atf}, \cite{Visser:2021ucg}. The real Lorentzian metric is not actually in the set of allowable complex metrics. Instead the real Lorentzian metric is at the boundary of the set of allowable complex metrics \cite{Kontsevich:2021dmb}, \cite{Witten:2021nzp}. The expectation we have is that through a smooth complex deformation of the Lorentzian metric, any physical quantity computed for the Euclidean metric will be equal to the same physical quantity in the Lorentzian spacetime that lives on the boundary of the space of complex metrics. There are other ways to extract physical quantities of the $(1,0)$ theory, for instance by following the ideas in \cite{Andriolo:2021gen}. However, there again one needs to consider a complex deformation of the Lorentzian spacetime metric and if one then is interested in the supersymmetric theory, one would inevitably be led to again study nonchiral conformal Killing spinors in the $(1,1)$ superconformal theory in Euclidean signature. In the end any method to compute a physical quantity should be equivalent with the method of holomorphic factorization starting from the $(1,1)$ superconformal theory in Euclidean signature.

Next we describe the conformal field theories in six dimensions. For the $(2,0)$ theories in Minkowski space we have an $OSp(8|2)$ superconformal group whose bosonic subgroup is the $SO(2,6) \times SO(5)$ conformal symmetry times the R-symmetry $SO(5) = Sp(2)$. There are $16$ Poincare supersymmetries that we parametrize by six-dimensional symplectic Majorana-anti Weyl spinors $\rho$ and $16$ special conformal supersymmetries that we parametrize by symplectic Majorana-Weyl spinors $\eta$. The general solution to the conformal Killing spinor equation is $\eps = \rho + \Gamma_M \eta x^M$ where $x^M$ are the six Lorentzian coordinates. 

For the six-dimensional $(1,0)$ theories in Minkowski space we have an $OSp(8|1)$ superconformal group whose bosonic subgroup is the same $SO(2,6)$ conformal symmetry but where the R-symmetry is reduced down to $Sp(1) = SU(2)$. The most general solution to the conformal Killing spinor equation is $\eps_I = \rho_I + \Gamma_M \eta_I x^M$ where $I,J,... = 1,2$ are indices in the fundamental representation of $SU(2)$ and $\eps_I$ are four-component Weyl spinors for each value $I=1,2$, which are subject to a symplectic Majorana condition. So we have $8$ real Poincare supersymmetries parametrized by $\rho_I$ and $8$ special conformal supersymmetries parametrized by $\eta_I$. When we reduce the R-symmetry, the tensor multiplet gets smaller, and to recover the field content of the $(2,0)$ tensor multiplet, we need to add two hypermultiplets whose fields we label by an index $A = 1,2$. We may also generalize that and consider an arbitrary number $N_f$ of hypers for which $A = 1,2,...,N_f$. These may live in arbitrary representations of the gauge group. But for convenience we will assume that our hypers live in the adjoint representation.

We will consider these theories on curved Lorentzian six-manifolds. The number of supersymmetries may then get reduced, but the field content remains the same. We shall define the theory to be either a $(2,0)$ theory or a $(1,0)$ theory according to the field content of their supermultiplets, where the $(1,0)$ theory has $2n$ real supercharges, and the $(2,0)$ theory has $4n$ real supercharges. Here $n$ is the number of solutions to the conformal Killing spinor equation for one complex Weyl spinor and hence $n$ is determined by the geometry of the six-manifold. Its maximal value is $n=8$, which is the number of complex solutions in Minkowski space where the $(1,0)$ and $(2,0)$ theories have $2n = 16$ and $4n = 32$ real supercharges, respectively.

For the problem of finding the nonabelian generalization of the $(1,0)$ theories, our strategy we will be to first reformulate the abelian tensor and hypermultiplets in loop space. In loop space the abelian two-form gauge potential $B_{MN}$ becomes a one-form defined as
\bea
A(C) &=& \int ds B_{MN}(C(s)) \dot{C}^M(s) \delta C^N(s)
\eea
Here $\delta C^M(s)$ can be thought of as a one-form differential in the infinite-dimensional loop space. Then the integral over $s$ corresponds to an `index contraction' of the continuous index $s$ in a reparametrization invariant manner. With a one-form gauge potential, we can construct a nonabelian gauge covariant derivative and obtain a nonabelian formulation. We will be making the following assumption regarding the nonabelian fields in loop space. We will assume that they are Lie algebra valued, and in the adjoint representation. Hence, concretely, we will assume that the nonabelian gauge field is given by $A(C) = A^a(C) T_a$ where $T_a$ are just the usual hermitian generators of the Lie algebra of the gauge group.  

We will not attempt to find a nonabelian generalization of $B_{MN}$. One such nonabelian construction can been found \cite{Baez:2004in}, \cite{Schreiber:2005ff} that uses both a nonabelian two-form $B_{MN}$ and a nonabelian one-form $A_{M}$ with field strength $F_{MN}$. By assuming both $B_{MN}$ and $A_M$ are in the adjoint representation of the same gauge group, which is what one would expect in a supersymmetric nonabelian tensor multiplet theory, then reparametrization invariance of a nonabelian Wilson surface requires us to impose the condition
\ben
B_{MN} &=& F_{MN}\label{fake}
\een
The field strength of $B_{MN}$ when this condition is satisfied is given by
\bea
H_{MNP} &=& 3D_{[M} B_{NP]}
\eea
The condition (\ref{fake}) then implies that 
\bea
H_{MNP} &=& 0
\eea
by using the Bianchi identity $3 D_{[M} F_{NP]} = 0$. We do not want to have such a strong condition on $H_{MNP}$ because that is incompatible with the abelian tensor multiplet where $H_{MNP}$ can be nonzero.

We will focus our attention on the loop space formulation itself without worrying about how the loop space fields might be explicitly realized in terms of local fields. To show the existence of a loop space reformulation is already a nontrivial problem, especially when the Lorentzian six-manifold is curved. 

We will rewrite the supersymmetry variations in a cohomological form by absorbing the supersymmetry parameter into the fermionic fields. This makes all fields, bosonic as well as fermionic, become antisymmetric tensor fields or scalar fields. Let us consider the supersymmetry variation of the abelian two-form gauge potential
\bea
\delta B_{MN} &=& i \bar\eps^I \Gamma_{MN} \lambda_I
\eea
This induces a corresponding supersymmetry variation of the one-form in loop space,
\bea
\delta A(C) = \int ds \delta B_{MN} \dot{C}^M \delta C^N = i \int ds \bar\eps^I \Gamma_{MN} \lambda_I \dot{C}^M \delta C^N
\eea
Let us assume that we want a loop space formulation in Minkowski space where the supersymmetry parameter is given by $\eps_I = \rho_I + \Gamma_M \eta x^M$. It depends in general on spacetime unless we would be interested in only preserving the Poincare supercharges, which are those parametrized by constant spinors $\rho_I$. One may be interested in only preserving the Poincare supercharges if one wants to perform dimensional reduction to a five-dimensional super Yang-Mills theory. But since our goal is to describe the six-dimensional theory, we should not break the special conformal supersymmetries, which are spacetime dependent. In curved spacetime the situation gets only worse as there we may not have any constant parameter $\eps_I$ such that  it can be brought outside the integral over the loop. Preserving all these supersymmetries is essential in any formulation of these theories because the superconformal symmetry is a defining property of these theories. These considerations leads us to include $\bar\eps^I$ into the definition of a new fermionic field
\bea
\Psi_{MN} &=& \bar\eps^I \Gamma_{MN} \lambda_I
\eea
in a cohomological reformulation. Then we may define a corresponding fermionic field in loop space as 
\bea
\Psi(C) &=& \int ds \Psi_{MN} \dot{C}^M \delta C^N
\eea
With these definitions, our supersymmetry variation becomes 
\ben
\delta A(C) &=& i \Psi(C)\label{L}
\een
In the cohomological reformulation, the parameters $\eps_I$ reappear in the form of bilinear combinations $V^M$, $\Theta_{MNP,IJ}$ and $R^I{}_J$. A different choice of a pair of spinors $\eps_I$ can lead to a different set of bilinear fields. We may get a different vector field $V^M$ and so on. So what we do, is that we pick two solutions $\eps_I$ to the conformal Killing spinor equation. Then we map the fields to their cohomological form, and then to the loop space. When we map the fermionic fields to their cohomological form, it is important that we do not lose any components. We shall have an equal number of fermionic field components in either formulation. To make sure that we have collected all components, we need to make sure that the map between fermionic spinor fields to fermionic tensor fields is invertible. We show that these maps indeed are invertible, for both the tensor and  hypermultiplet fermionic fields. (The explicit forms of these inverse maps are presented in equation (\ref{invtensor}) for the tensor multiplet fermions and in the equations (\ref{invhyper1}), (\ref{invhyper2}) and (\ref{invhyper3}) for the hypermultiplet fermions.) 

It is a different question how many supersymmetries we preserve. We should preserve all the global symmetries, and in particular we should preserve all the supersymmeries, which are present in the original abelian six-dimensional theory for a given Lorentzian six-manifold. If we do not preserve all the global symmetries, then we have got a different theory because the global symmetries are a defining property of a theory. The vector field $V^M$ does not appear in the loop space gauge field $A(C)$, nor in the fermionic field $\Psi(C)$ that we introduced above. However, it does appear in our definition of our loop space scalar field,
\bea
\Phi(C) &=& \int ds \dot{C}^M(s) V_M(C(s)) \phi(C(s))
\eea
This definition leads to the desired property that its Lie derivative along $V$ is given by  
\bea
\L_V \Phi(C) &=& \int ds \dot{C}^M(s) V_M(C(s)) \(\L_V \phi(C(s)) + \frac{\Omega}{3} \phi\)
\eea
This is the desired property of this Lie derivative because in the cohomological formulation the exact same combination appears in the supersymmetry variation of the cohomological fermionic field $\Psi = \bar\eps^I \lambda_I$ as
\ben
\delta \Psi &=& \L_V \phi + \frac{\Omega}{3} \phi\label{CH}
\een
This means that by defining the loop space field $\Phi(C)$ as above and a corresponding fermionic loop space field as
\bea
\Psi(C) &=& \int ds \dot{C}^M(s) V_M(C(s)) \Psi(C(s))
\eea
we can uplift the variation into loop space where it becomes
\bea
\delta \Psi(C) &=& \L_V \Phi(C)
\eea
The fact that we can uplift the supersymmetry variation (\ref{CH}) to loop space is highly nontrivial. We could imagine having defined our loop space fields differently, as
\bea
\Phi(C) &=& \int ds e(s) \phi(C(s))\cr
\Psi(C) &=& \int ds e(s) \Psi(C(s))
\eea
where we introduce a one-bein $e(s) = \sqrt{g_{MN}(C(s)) \dot{C}^M(s) \dot{C}^N(s)}$ along the loop. These definitions are reparametrization invariant, and they could also seem to be suggested by the way that the Wilson surface is defined in \cite{Gustavsson:2004gj} and the fact that a Wilson surface would be a Wilson line or Wilson loop in loop space. But with such definitions we would not be able to uplift the supersymmetry variation (\ref{CH}) to loop space. The Lie derivative $\L_V \Phi(C)$ would come out differently, and we can not uplift the second term $\frac{\Omega}{3} \phi$ to loop space as a separate term because $\Omega$ is in general not a constant that we could take outside the integral around the loop. Another objection to using such a definition of $\Phi(C)$ comes from the fact that it can not be used to get the right Wilson surface after all, which uses the measure of the surface rather than the measure along a foliation loop inside that surface.

Now let us discuss the fact that our definition of $\Phi(C)$ depends on $V^M$ and hence on the supersymmetry parameters that we pick. While this is true, it does not mean that the Lagrangian, if it can be constructed in loop space at the end of the day, would not be invariant under all supersymmetries corresponding to all different choices of supersymmetry parameters. 

More concretely we can see the same thing happens for the supersymmetry variations in the cohomological formulation. There again the cohomological fermionic fields depend on the choice of $\eps_I$ and hence in an indirect way also on the vector field $V^M$, much like our loop space fields $\Phi(C)$ and $\Psi(C)$ depend on $V^M$. But for the cohomological reformulation, we know that the Lagrangian has the same supersymmetries as the original Lagrangian because the two Lagrangians are related by a field redefinition, which does not break any global symmetries.

We do not have off-shell supersymmetry for the abelian tensor multiplet. In the closure relation on the tensor gauge potential $B_{MN}$, we find a term that is proportional to the antiselfdual part of the field strength,
\bea
\delta^2 B_{MN} &=& ... - i V^P H^-_{PMN} 
\eea
assuming a commuting supersymmetry parameter. Here the antiselfdual part is defined as
\bea
H^-_{MNP} &=& \frac{1}{2} \(H_{MNP} - \frac{1}{6} \eps_{MNP}{}^{RST} H_{RST}\)
\eea
and the field strength is defined as
\bea
H_{MNP} &= &\partial_M B_{NP} + \partial_N B_{PM} + \partial_P B_{MN}
\eea
When we reformulate the supersymmetry variations in a cohomological form, the antiselfdual term in the closure relation above reappears in the supersymmetry variation of the fermionic tensor fields $\Psi_{MN}$ as 
\bea
\delta \Psi_{MN} &=& ... + V^P H_{PMN}^-
\eea
This antiselfdual term contracted with $V$ can not be nicely expressed in term of our loop space fields. So we will proceed by assuming that this term is absent in the supersymmetry variation $\delta \Psi_{MN}$, which we can do if we assume that the fields live on the constraint surface where $H_{MNP} - \frac{1}{6} \eps_{MNP}{}^{RST} H_{RST} = 0$. On the constraint surface, the cohomological form is in one-to-one correspondence with the usual spinorial supersymmetry variations even if we drop the term $V^P H_{PMN}^-$, and also, on the constraint surface, we also have an invertible map from loop space back to the cohomological formulation, at least for the case of an abelian gauge group. The selfduality constraint is something that we need to impose by hand on top of the field equations. We notice that it will be sufficient to just imposing the constraints $V^P H^-_{PMN} = 0$. The number of selfduality constraints are $\frac{6 \times 5 \times 4}{3!}/2 = 10$. The number of contracted selfduality constraints with $V$ are $\frac{5 \times 4}{2!} = 10$. So they give us equally many conditions, and they are equivalent, as one may see explicitly by for example using the vielbein formulation.

Let us now present how the map from loop space back to the cohomological formulation  may be constructed, for the abelian case. Let us return to our sample supersymmetry variation above. The loop space is huge. So despite the variation 
\bea
\delta A(C) &=& i \Psi(C)
\eea
looks very simple, it contains a lot of information, since it holds for each shape of the loop $C$. For an infinitesimally small loop centered at a spacetime point $x$, we may define a two-form field $B_{MN}(x)$ implicitly through the relation
\bea
A(C) &=& B_{MN}(x) \delta \sigma^{MN}(C) 
\eea
where 
\bea
\delta \sigma^{MN}(C) &=& \int ds \frac{\partial C^M(s)}{ds} \delta C^N(s)
\eea
We notice that 
\bea
\sigma^{MN}(C) &=& \frac{1}{2} \int ds \(\frac{dC^M}{ds} C^N - \frac{dC^N}{ds} C^M\)
\eea
corresponds to the area inside $C$ and then $\delta \sigma^{MN}$ is describing how this area changes as we vary the loop infinitesimally. We can make this explicit by taking a circular loop in the $1,2$ plane centered at some point $x^M = (x^1,x^2,x^i)$ for $i = 1,2,3,4$,
\bea
C^1(s) &=& x^1 + r \cos s\cr
C^2(s) &=& x^2 + r \sin s\cr
C^i(s) &=& x^i
\eea
Then $\sigma^{12}(C) = - \pi r^2$ and $\delta \sigma^{12}(C) = - 2 \pi r d r$ if we make a homogeneous variation of the loop such that the new loop is located at the radius $r + dr$ and then 
\bea
A(C) &=& - B_{12}(x) 2 \pi r dr
\eea
from which we can extract $B_{12}(x)$ if we know $A(C)$, the loop $C$ and its variation $\delta C$. Thus we can extract $B_{MN}(x)$ from $A(C)$ for the abelian case. It would be very interesting if we can apply the same procedure to the nonabelian loop space gauge potential to extract a corresponding nonabelian two-form gauge potential. We leave this as a question for the future. 

Returning to the abelian case, a corresponding definition can be made for $\Psi(C)$,
\bea
\Psi(C) &=& \Psi_{MN}(x) \delta \sigma^{MN}(C)
\eea
This shows that we can recover the cohomological supersymmetry variations from the loop space supersymmetry variations, at least for the abelian case.

In order to find the nonabelian loop space theory, we will seek guidance from the mini loop space, which consists of those loops that wrap around a compact circle direction that is generated by a spacelike Killing vector field $v^M$. These loops are therefore the orbits of the Killing vector field. We reduce our abelian loop space supersymmetry variations down to the mini loop space by picking the zero mode of the spacetime field that we integrate around the orbit in the mini loop space. For the two-form gauge potential, we define the corresponding mini loop space one-form gauge field as
\bea
A(x) &=& B_{MN}(x) v^M(x) dx^N
\eea
that we shall combine with the constraint equation
\bea
\L_v A(x) &=& 0
\eea
The definitions of all our fields in the mini loop space are given by the equations (\ref{vSYMtomini}) and (\ref{hSYMtomini}) for the tensor and hypermultiplets respectively. Our next step is to map the supersymmetry variations of 5d super Yang-Mills from the cohomological form to the mini loop space form. In this way we find the nonabelian generalization in the mini loop space, from the known nonabelian supersymmetry variations of 5d super Yang-Mills. We present these results in sections \ref{mini-non-tensor} and \ref{mini-non-hyper} for the vector and hypermultiplets respectively. From these derived results in the mini loop space for the nonabelian generalization, we are then finally able to conjecture the corresponding nonabelian supersymmetry variations in the full loop space for the tensor and hypermultiplets.

\section{The six-dimensional superconformal theories}\label{section3}
In the following two subsection we summarize the supersymmetry variations of the abelian six-dimensional superconformal theories with $(2,0)$ and $(1,0)$ supersymmetries respectively. The nonabelian generalizations in six dimensions are not known.  

\subsection{The $(2,0)$ tensor multiplet}
The Poincare supercharges of the $(2,0)$ tensor multiplet 
on Minkowski spacetime $\mb{R}^{1,5}$ 
transform in the representation $(4,4)$ of the $SO(1,5) \times SO(5)_R$ Lorentz group times the R-symmetry group. The anticommutator of two supercharges therefore transforms in the symmetric representation\footnote{Here subscripts $s$ and $a$ stand for symmetric and antisymmetric respectively.}
\bea
((4,4) \otimes (4,4))_s &=& (6_a\oplus 10_s,1_a \oplus 5_a \oplus 10_s)_s\cr
&=& (6,1) \oplus (6,5) \oplus (10,10)
\eea
The three terms on the second line correspond to the momentum $P_M$ and to two central charges $Z_M^A$ and $W_{MNP}^{AB}$ of a selfdual string and of a three-dimensional brane respectively, associated with the intersection of an M2 brane and an M5 brane respectively. 

We can also put the (2,0) tensor multiplet on a curved Lorentzian six-manifold that has a conformal Killing spinor, which is a spinor $\eps$ that satisfies 
\ben
\nabla_M \eps &=& \Gamma_M \eta\label{CKE}
\een
for some other spinor $\eta$. Both $\eps$ and $\eta$ are transforming in the four-component spinor representation of the $SO(5)$ R-symmetry. Hence, if there are $n$ complex spinorial solutions to the conformal Killing spinor equation in 6d when we do not attach any R-symmetry index, then there will be in total $4n$ real supercharges in the (2,0)-theory. We note that on Minkowski spacetime $n=8$ and we have $32$ real supercharges, whereof $16$ are conformal supercharges and the other $16$ are Poincare supercharges. For a generic six-manifold that admits $n$ complex conformal Killing spinors, we have for each of the $4n$ solutions $\eps$ to (\ref{CKE}), the following supersymmetry variations of the $(2,0)$ tensor multiplet 
\bea
\delta \phi^A &=& i \bar\eps \Gamma^A \psi\cr
\delta B_{MN} &=& i \bar\eps \Gamma_{MN} \psi\cr
\delta \psi &=& \frac{1}{12} \Gamma^{MNP} \eps H_{MNP} + \Gamma^M \Gamma^A \eps \partial_M \phi^A - 4 \Gamma^A \eta \phi^A
\eea
We use 11d gamma matrices that we split into spacetime and R-symmetry components $\Gamma^M$ and $\Gamma^A$ respectively, such that
\bea
\{\Gamma_M,\Gamma_N\} &=& 2 g_{MN}\cr
\{\Gamma^A,\Gamma^B\} &=& 2 \delta^{AB}\cr
\{\Gamma^M,\Gamma^A\} &=& 0
\eea
where $g_{MN}$ denotes the metric tensor on the Lorentzian six-manifold. We impose the 11d Majorana condition
\bea
\bar\eps &=& \eps^T C_{11d}
\eea
where the Dirac conjugate is defined as $\bar\eps =  \eps^{\dag} \Gamma^0$. We define the product of eleven 11d gamma matrices to be the unit $32\times 32$ matrix
\bea
\Gamma^{012345} \Gamma^{\h{1}\h{2}\h{3}\h{4}\h{5}} &=& 1
\eea
in flat Minkowski and in a general curved spacetime the unit matrix on the right-hand side shall be multiplied with $\sqrt{-g}$. We impose the 6d Weyl projections
\bea
\Gamma \eps &=& - \eps\cr
\Gamma \psi &=& \psi\label{6dWeyl}
\eea
on the supersymmetry parameter $\eps$ and the spinor field $\psi$, where $\Gamma = \Gamma^{012345}$ in flat Minkowski and in general curved spacetime we need to also divide the right-hand side by $\sqrt{-g}$ so as to have $\Gamma^2 = 1$. We also have 
\bea
\Gamma \eta &=& \eta
\eea
If we represent the 11d gamma matrices as
\bea
\Gamma^M &=& \Gamma^M \otimes 1\cr
\Gamma^A &=& \Gamma \otimes \gamma^A\
\eea
where with a slight abuse of notation we recycle $\Gamma^M$, then the supersymmetry variations become % read
\bea
\delta \phi^A &=& i \bar\eps \gamma^A \psi\cr
\delta B_{MN} &=& i \bar\eps  \Gamma_{MN} \psi\cr
\delta \psi &=& \frac{1}{12} \Gamma^{MNP} \eps H_{MNP} - \Gamma^M \gamma^A \eps \partial_M \phi^A - 4 \gamma^A \eta \phi^A
\eea

\subsection{The $(1,0)$ tensor and hypermultiplets} 
The Poincare supercharges of the $(1,0)$ tensor multiplet transform in the representation $(4,2)$ of the $SO(1,5) \times SU(2)_R$ Lorentz times the R-symmetry group. The anticommutator of two supercharges therefore transform in the symmetric representation 
\bea
((4,2) \otimes (4,2))_s &=& (6_a\oplus 10_s,1_a \oplus 3_s)_s\cr
&=& (6,1) \oplus (10,3) 
\eea
The terms on the second line correspond to the momentum $P_M$ and one central charge $Z_{MNP}^{IJ}$ that is symmetric in $IJ$ and selfdual in $MNP$. 

We may obtain $SU(2)_R$ by breaking the $SO(5)$ R-symmetry by projecting the supersymmetry parameter 
\bea
\gamma^5 \eps &=& - \eps\cr
\bar\eps \gamma^5 &=& - \bar\eps\cr
\gamma^5 \eta &=& - \eta\label{RWeyl}
\eea
Then the $(2,0)$ tensor multiplet will split into one $(1,0)$ tensor multiplet whose fermions will be denoted $\lambda$ and one $(1,0)$ hypermultiplet whose fermions will be denoted $\psi_{(1,0)}$. These fermions are the two Weyl components of the original $(2,0)$ fermion and thus satisfy 
\bea
\gamma^5 \lambda &=& - \lambda\cr
\gamma^5 \psi_{(1,0)} &=& \psi_{(1,0)}
\eea
Since we will not discuss the $(2,0)$ tensor multiplet any further, we will drop the subscript on $\psi_{(1,0)}$ and write $\psi$ for simplicity. The supersymmetry variations are
\bea
\delta \phi &=& - i \bar\eps \lambda\cr
\delta B_{MN} &=& i \bar\eps \Gamma_{MN} \lambda\cr
\delta \lambda &=& \frac{1}{12} \Gamma^{MNP} \eps H_{MNP} + \Gamma^M \eps \partial_M \phi + 4 \eta \phi
\eea
and 
\bea
\delta \phi^i &=& i \bar\eps \gamma^i \psi\cr
\delta \psi &=& - \Gamma^M \gamma^i \eps \partial_M \phi^i - 4 \gamma^i \eta \phi^i
\eea
Here the five scalar fields split into one tensor multiplet scalar field $\phi = \phi^5$ and four hypermultiplet scalar fields $\phi^i$ for $i = 1,2,3,4$. 

It is custom to display the R-symmetry spinor indices $I,J,...$ explicitly. Then we have
\ben
\delta \phi &=& - i \bar\eps^I \lambda_I\cr
\delta B_{MN} &=& i \bar\eps^I \Gamma_{MN} \lambda_I\cr
\delta \lambda_I &=& \frac{1}{12} \Gamma^{MNP} \eps_I H_{MNP} + \Gamma^M \eps_I \partial_M \phi + 4 \eta_I \phi\label{getback}
\een
and 
\bea
\delta \phi^i &=& i \bar\eps^I (\sigma^i)_{IA} \psi^A\cr
\delta \psi^A &=& - \Gamma^M (\sigma^i)^{AI} \eps_I \partial_M \phi^i - 4 (\sigma^i)^{AI} \eta_I \phi^i
\eea
where we have introduced the half gamma matrices $(\sigma^i)_{IA}$ and $(\sigma^i)^{AI}$ that sit in the full gamma matrices as off-diagonal blocks,
\bea
\gamma^i &=& \(\begin{matrix}
0 & (\sigma^i)^{AJ}\\
(\sigma^i)_{IB} & 0
\end{matrix}\)
\eea
We also have
\bea
\gamma^5 &=& \(\begin{matrix}
\delta^A_B & 0\\
0 & - \delta_I^J
\end{matrix}\)
\eea

We may further define 
\bea
q^{AI} &=& \phi^i (\sigma^i)^{AI}
\eea
and, by using the identity,
\bea
(\sigma^i)^{AI} (\sigma^i)_{JB} &=& 2 \delta^A_B \delta^I_J
\eea
we get 
\bea
\delta q^{AI} &=& 2 i \bar\eps^I \psi^A\cr
\delta \psi^A &=& - \Gamma^M \eps_I \partial_M q^{AI} - 4 \eta_I q^{AI}
\eea

For $SU(2)$ R-symmetry, if our Lorentzian six-manifold admits $n$ complex conformal Killing spinors, then we will have in this $(1,0)$ theory $2n$ real supercharges. For Minkowski space, $n=8$ and so we have $16$ real supercharges, half of which are conformal supercharges, and the other half are Poincare supercharges.

The rules for charge conjugation of spinors in 6d are 
\bea
C_6^T &=& C_6\cr
\Gamma_M^T &=& - C_6 \Gamma_M C_6^{-1}
\eea
By using these rules, one can show that
\bea
\Theta^{MNP}_{IJ} &=& \bar\eps_I \Gamma^{MNP} \eps_J
\eea
is symmetric in $I$ and $J$ by also using the Majorana condition
\bea
\bar\eps_I &=& \eps_I^T C_6\label{313}
\eea
but this relation needs to be carefully interpreted. It is not a purely six-dimensional Majorana condition since the Dirac conjugated spinor is defined as
\bea
\bar\eps^I_{\dot\beta} &=& \(\eps^{\alpha}_I\)^* (\Gamma^0)^{\alpha}{}_{\dot\beta}
\eea
with the index $I$ upstairs. Here we also display the two four-component spacetime Weyl and anti-Weyl spinor indices $\alpha$ and $\dot\alpha$ explicitly. Then we see that the Majorana condition requires the use of the antisymmetric tensor $\ep^{JI}$, thus exhibiting the symplectic nature of the Majorana condition,
\bea
\bar\eps^I_{\dot\alpha} &=& \eps_J^{\beta} \ep^{JI} C_{\beta\dot\alpha}
\eea
Then %if 
lowering the index $I$ on both sides by $\ep_{IJ}$ using $\ep^{IJ} \ep_{JK} = \delta^I_K$, we recover (\ref{313}),
\ben
\bar\eps^I_{\dot\alpha} \ep_{IJ} := \bar\eps_{J\dot\alpha} = \eps_J^{\beta} C_{\beta\dot\alpha}\label{maj2}
\een
We have a lightlike conformal Killing vector  
\bea
V^M &=& \bar\eps^I \Gamma^M \eps_I
\eea
By using (\ref{maj2}) we find that $\bar\eps_I \Gamma^M \eps_J$ is antisymmetric in $I$ and $J$ and then
\bea
\bar\eps_I \Gamma^M \eps_J &=& - \frac{1}{2} \ep_{IJ} V^M
\eea
We have the Fierz identity 
\ben
\eps_I \bar\eps_J &=& \frac{1}{8} \ep_{IJ} V^M \Gamma_M - \frac{1}{24} \Theta^{MNP}_{IJ} \Gamma_{MNP}\label{F1}
\een
which can be used to show that $V^M V_M = 0$ and $\Theta_{MNP}{}^I{}_J V^P = 0$. By using the relations
\bea
\nabla_M V^M &=& 12 \bar\eps^J \eta_J\cr
\nabla_{[M} V_{N]} &=& - 2 \bar\eps^J \Gamma_{MN} \eta_J
\eea
we can obtain another useful Fierz identity,
\bea
\eta_J \bar\eps^J &=& \frac{1}{48} \nabla_M V^M + \frac{1}{16} \Gamma^{MN} \nabla_M V_N
\eea

\section{The lightlike vector $U^M$}\label{section4}
In this section we show that there exists a unique lightlike vector 
\bea
\frac{U^M}{\N}
\eea
where $\N$ is a nowhere vanishing normalization factor,  
\bea
\N &=& V^M U_M
\eea
and $\L_V U^M = [V,U]^M = 0$. 

By applying the Fierz identity (\ref{F1}) on $\Gamma_M \eps_I V^M$ using that $\Gamma_M \Gamma^{RST} \Gamma^M = 0$, we get
\bea
\Gamma_M \eps_I \bar\eps^J \Gamma^M \eps_J = - \frac{1}{2} \Gamma_M \eps_I V^M
\eea
and therefore 
\bea
\Gamma_M \eps_I V^M &=& 0\label{confused}
\eea
For a lightlike $V^M$ this is a Weyl projection. This is best seen by switching to flat tangent space indices $A,B,...$ (not to be confused with SO(5) R-symmetry indices) by using the vielbein $e^A_M$. We can apply a local tangent space Lorentz transformation to get $V^A$ on the form 
\ben
V^A &=& \frac{V}{\sqrt{2}} (1,0,0,0,0,1)\label{VA}
\een
for some function $V$ that is fixed by the conformal Killing vector property up to a constant factor. We shall assume that $V$ is nowhere vanishing. Then the Weyl projection (\ref{confused}) becomes 
\bea
\Gamma^{05} \eps_I &=& - \eps_I
\eea
To express the projection operator in a covariant language, we need to introduce another lightlike vector field $U_M$ that is linearly independent of $V_M$. This implies that $g_{MN} V^M U^N$ is nowhere vanishing. We may then without loss of generality assume that $g_{MN} V^M U^N = 1$ by an appropriate local rescaling and possible also a change of the sign of $U^M$. Just like $V^M$ is Weyl invariant, we will make the assignment that $U^M$ is Weyl invariant. This is a convenient assignment since then the commutator $[V,U]^M$ will also be Weyl invariant. However, if we fix $g_{MN} V^M U^N = 1$ then we break the conformal invariance. Let us accordingly denote the metric that we are using here as $\t{g}_{MN}$ just to indicate that it is with respect to this metric that we normalize $U^M$ such that $\t{g}_{MN} V^M U^N = 1$ and if we make a subsequent Weyl transformation $\t{g}_{MN} \rightarrow g_{MN} = e^{2\sigma} \t{g}_{MN}$ then the normalization will change such that $g_{MN} V^M U^N = e^{2\sigma} > 0$ and this is how we will restore the Weyl invariance shortly, by allowing for an arbitrary non-negative local normalization. But for now let us proceed with this specific metric $\t{g}_{MN}$, and let us make the following ansatz for the Weyl projection operators,
\bea
P_- &=& \frac{1}{2} U_M V_N \Gamma^M \Gamma^N\cr
P_+ &=& \frac{1}{2} V_M U_M \Gamma^M \Gamma^N
\eea
in this metric. One can then show that 
\bea
P_+^2 &=& P_+\cr
P_-^2 &=& P_-\cr
P_+ P_- &=& 0\cr
P_+ + P_- &=& 1
\eea
which are all nice properties that we shall have for the Weyl projection operators. But we also need them to be hermitian. To understand the implications of that, we switch to flat tangent space indices that we decompose as $A = (0,a)$ where $a=1,...,5$ and then we expand 
\bea
P_- = \frac{1}{2} \(-U_0 V_0 + \(V_a U_0 - U_a V_0\) \Gamma^{0a} + U_a V_b \Gamma^{ab}\)
\eea
By demanding this to be hermitian we get the conditions 
\bea
U_a V_b - U_b V_a &=& 0
\eea
With our choice of $V^A$ as in (\ref{VA}) these conditions determine $U^A$ uniquely as
\ben
U^A &=& \frac{1}{V\sqrt{2}} \(-1,0,0,0,0,1\)\label{UA}
\een
From (\ref{UA}) we can also see that $U_M \Gamma^M$ gives a Weyl projection of opposite chirality compared to $V_M \Gamma^M$. As such it should change at most by a local rescaling as we move along the vector field $V^M$. Hence we shall have that 
\bea
\L_V \(U_M \Gamma^M\) &\sim & U_M \Gamma^M
\eea
Expanding the Lie derivative, we get
\bea
\(\frac{\t\Omega_V}{6} U_M + [V,U]_M\) \Gamma^M &\sim & U_M \Gamma^M
\eea
From this we conclude that
\bea
[V,U]^M &=& c U^M
\eea
for some scalar function $c$. Let us note that for an arbitrary vector field $X^M$ we have by just using the conformal vector equation for $V^M$ that  
\bea
\L_V \(V^M X_M\) &=& \frac{\t\Omega}{3} V^M X_M + V^M [V,X]_M
\eea
where $\t\Omega = \nabla_M V^M$ as computed with respect to the metric $\t{g}_{MN}$. If we apply this relation to $X_M = \t{g}_{MN} U^N$ then we get
\bea
\t{g}_{MN} V^M [V,U]^N &=& - \frac{\t\Omega}{3} 
\eea
and thus 
\bea
c &=& - \frac{\t\Omega}{3}
\eea
is completely fixed by $V^M$ and the metric $\t{g}_{MN}$. Let us next examine what a local rescaling of $U^M$ (not a Weyl transformation) does. If we put $U^M = e^q \t{U}^M$ for some function $q$, then we get
\bea
[V,\t{U}]^M &=& \(- \L_V q - \frac{\t\Omega}{3}\) \t{U}^M\cr
\t{g}_{MN} V^M \t{U}^M &=& e^{-q}
\eea
Thus by choosing $q$ so that 
\bea
\L_V q &=& - \frac{\t{\Omega}}{3}
\eea
then we get
\bea
[V,\t{U}]^M &=& 0
\eea
We notice that $q$ is completely determined by $V^M$ and $\t{g}_{MN}$ up to an additive integration constant that here need not be a constant. Namely, suppose that we pick a local coordinate system where $\L_V = \partial_+$. Then 
\bea
\partial_+ q &=& - \frac{\t{\Omega}}{3}
\eea
that we can solve by making just one integration as
\bea
q &=& q_0(x^-,x^i) - \int_0^{x^+} dx'^+ \frac{\t{\Omega}}{3}
\eea
The integration constant $q_0$ may thus be a arbitrary function in the transverse directions to $V^M$. Then $\t{U}^M$ is completely fixed by $V^M$ and $\t{g}_{MN}$ up to a non-vanishing multiplicative function $C = e^{-q_0}$ in the transverse space. Such a multiplicative function is a remaining freedom that can not be fixed by requiring that the commutator $[V,\t{U}]^M$ vanishes.

When we make a Weyl transformation of the metric $\t{g}_{MN} \rightarrow g_{MN} = e^{2\sigma} \t{g}_{MN}$ we get with respect to the new metric that
\bea
\N := g_{MN} V^M \t{U}^N = e^{2\sigma - q} > 0
\eea
Let us now drop the tilde on $\t{U}^M$ for notational simplicity. Then we can summarize our results by the equations
\bea
V^M U_M &=& \N > 0\cr
[V,U]^M &=& 0\cr
\nabla_M V_N + \nabla_N V_M &=& \frac{\Omega}{3} g_{MN}\cr
V^M V_M &=& 0\cr
U^M U_M &=& 0
\eea
We notice that $U^M$ is completely fixed by $V^M$ and the metric $g_{MN}$ up to a multiplicative function in the transverse space to $V^M$. It is true that $U^M$ was fixed with respect to the metric $\t{g}_{MN}$, but since $U^M$ is Weyl invariant it does not depend on which representative metric we choose, so we are free to make a Weyl transformation and use the metric $g_{MN}$ for which we have the normalization $V^M U_M = \N > 0$ to fix $U^M$.  
%Also we notice that $U^M$ is in general not a conformal Killing vector field. 

We may also note that $U^M/\N$ is unambiguous, since the ambiguous multiplicative factor cancels between the numerator and the denominator. 

In order to restore the Weyl invariance in the projection operators, these should be defined as
\bea
P_- &=& \frac{1}{2 \N} U_M V_N \Gamma^M \Gamma^N\cr
P_+ &=& \frac{1}{2 \N} V_M U_M \Gamma^M \Gamma^N
\eea
Here we can also see that what appears is the ratio $U^M/\N$ and not the ambiguous vector field $U^M$ itself. 

\subsection{Is $U$ a conformal Killing vector?}
We can always make a Weyl transformation such that $V^M$ is a Killing vector with respect to the Weyl transformed metric. Let us assume that we have such a metric where $\L_V g_{MN} = 0$. Then we notice that 
\bea
\L_V \(\L_U g_{MN}\) = \L_U \L_V g_{MN} = 0
\eea
This suggests, but does not completely prove, that 
\bea
\L_U g_{MN} &=& \frac{\Omega^{\vee}}{3} g_{MN}
\eea
where we define 
\bea
\Omega^{\vee} &=& \nabla_M U^M
\eea
If this is true, then $U$ would be a conformal Killing vector. We also need to have
\bea
\L_V \Omega^{\vee} &=& 0
\eea
But this we can actually prove. We start with noting the identity
\bea
\L_U g_{MN} &=& \nabla_M U_N + \nabla_N U_M
\eea
that does not assume any properties of $U^M$, but is just a consequence of $\nabla_P g_{MN} = 0$. Next,
\bea
\Omega^{\vee} &=& \frac{1}{2} g^{MN} \L_U g_{MN}
\eea
So we get
\bea
\L_V \Omega^{\vee} &=& \frac{1}{2} \L_V g^{MN} \L_U g_{MN} + \frac{1}{2} g^{MN} \L_V \L_U g_{MN}
\eea
Here the first term is zero by $\L_V g^{MN} = - \nabla^M V^N - \nabla^N V^M = 0$ and the second term is zero by $\L_V \L_U g_{MN} = \L_U \L_V g_{MN} = 0$, hence we have now proved that
\bea
\L_V \Omega^{\vee} &=& 0
\eea
Furthermore
\bea
V^M V^N \L_U g_{MN} = \L_U \(V^M V^N g_{MN}\) = 0
\eea
since $\L_U V^M = [U,V]^M = 0$. Similarly
\bea
U^M U^N \L_U g_{MN} = \L_U \(U^M U^N g_{MN}\) = 0
\eea
because $\L_U U^M = 0$ as an identity that does not assume anything of $U^M$. Let us summarize. We have been able to prove the following results,
\bea
V^M V^N \L_U g_{MN} &=& 0\cr
U^M U^N \L_U g_{MN} &=& 0\cr
\Omega^{\vee} &=& \frac{1}{2} g^{MN} \L_U g_{MN}\cr
\L_V \Omega^{\vee} &=& 0\cr
\L_V g_{MN} &=& 0\cr
\L_V U^M &=& 0\cr
\L_V \L_U g_{MN} &=& 0
\eea
All these relations still hold if we make a rescaling of $U^M$ such that 
\bea
U^M &\rightarrow & C U^M\cr
\L_V C &=& 0
\eea
We also note that all the relations above are consistent with, but do not prove at this point that we actually have, the conformal Killing vector equation
\bea
\L_U g_{MN} &=& \frac{\Omega^{\vee}}{3} g_{MN}
\eea
Quite the contrary, it is highly unlikely that we would be able to prove that $U$ is a conformal Killing vector from the assumptions we have made. But if the conformal Killing vector equation would be satisfied, then we would lose the freedom that we saw above to rescale $U^M$ since then $U$ is determined up to just a constant multiplicative factor. We notice that the scale ambiguity can also be removed if one forms the ratio $U^M / \N$. But this ratio can not be a conformal Killing vector since it is not Weyl invariant. 

As we will see, the uplift to loop space requires $U$ to be a conformal Killing vector. In particular we can see that for the uplift to loop space of the supersymmetry variation of a particular fermionic field in the hypermultiplet, that we call $\Up_M^{AI}$. We show how this uplift is performed in detail in appendix \ref{UpQAI}. 

But so far we have been unable to show that $U$ necessarily must be a conformal Killing vector, although we can see that things are set up in such a way that our equations are compatible with having $U$ as a conformal Killing spinor, and yet it is clear that we need something more, since otherwise it is unlikely that all the $21$ conformal Killing vector equations $\L_U g_{MN} \sim g_{MN}$ will be satisfied by $U$. 

In the next subsection we will present a way in which we can show that $U$ is a conformal Killing vector that commutes with $V$. But it requires us to study a certain class of nonchiral conformal Killing spinors. Nonchiral spinors correspond to $(1,1)$ supersymmetry. The existence of nonchiral spinors is a natural condition to impose on the geometry because there is no manifestly covariant Lagrangian for the $(1,0)$ abelian tensor multiplet that can be used for quantization. Instead one may start with a nonchiral Lagrangian with $(1,1)$ supersymmetry and perform holomorphic factorization \cite{Witten:1996hc}, \cite{Henningson:1999dm}. Also, if one wants to study the theory in Euclidean signature, then one is forced to consider nonchiral spinors and $(1,1)$ supersymmetry. For instance, in \cite{Bastianelli:2000hi} the classical $(2,2)$ superconformal Lagrangian on a Riemannian six-manifold was used in order to compute the Weyl anomaly for the chiral $(2,0)$ quantum theory.

\subsection{Quantum $(1,0)$ superconformal theory}
Let us first notice that for Einstein manifolds, we have 
\bea
\nabla_M \eps_I &=& \Gamma_M \eta_I\cr
\nabla_M \eta_I &=& - \frac{R}{120} \Gamma_M\eps_I
\eea
So for Einstein manifolds there is a second set of conformal Killing spinors $\eta_I$ that we can use to form a lightlike conformal Killing vector as
\bea
U_M &=& \bar\eta^I \Gamma_M \eta_I
\eea
Moreover, since $\eta_I$ also has the opposite 6d Weyl chirality compared to $\eps_I$, it is clear that $U$ and $V$ are two distinct conformal Killing vectors.

In the appendix \ref{ConfEin} we show that six-manifolds that admit at least two nonchiral conformal Killing spinors with a certain normalization condition, are conformally equivalent to Einstein manifolds. Hence by a suitable Weyl transformation that brings the manifold into an Einstein manifold, we can construct a conformal Killing vector $U$ as the Dirac current of $\eta_I$ as above. The fact that $U$ is a conformal Killing vector does not change when we make a Weyl transformation. However, since $\eta_I$ has a rather complicated transformation under Weyl transformations, as
\ben
\eps & \rightarrow & e^{\frac{\sigma}{2}} \eps\cr
\eta &\rightarrow & e^{- \frac{\sigma}{2}} \(\eta + \frac{1}{2} \Gamma^M \eps \partial_M \sigma\)
\een
$U$ will in general not be the Dirac current of $\eta_I$ after a Weyl transformation. Moreover, in the same appendix we show that $[U,V] = 0$ and that $\Gamma^{MN} U_M V_N$ is hermitian. This shows that the $U$ that we construct as the Dirac current of $\eta_I$ for an Einstein manifold, must be the same $U$ (for some particular choice of normalization factor) as we constructed previously as a vector field to define Weyl projection operators $P_{\pm}$ in a covariant way in section \ref{section4}.

\section{Dimensional reduction}\label{section5}
For the nonabelian generalization, we need to perform a dimensional reduction down to five dimensions where we have a five-dimensional super Yang-Mills theory whose nonabelian structure is well-known. We will perform this dimensional reduction along a spatial Killing vector field $v$. Hence we assume that such a Killing vector exists here. But it shall be noted that this assumption is made only in order to find the nonabelian generalization. The nonabelian structure is something rather different from the geometric structure that we can see already in the abelian theories. We do not expect the nonabelian structure will depend in any crucial way on the geometric structure. Concretely, if we have a certain commutator term in Minkowski space, then the corresponding commutator term should also appear on a curved spacetime, and vice versa. 

There are many different ways in which we may formulate five-dimensional super Yang-Mills theory. Here we will present the formulation that preserves the six-dimensional covariance but where we constrain the fields to have vanishing Lie derivatives along $v$. We will then translate this formulation into the corresponding cohomological formulation where the fermionic spinor fields are replaced with fermionic tensor and scalar fields.

\subsection{The fermionic spinor field formulation}
We may formulate 5d super-Yang-Mills in a 6d covariant way. We assume that our Lorentzian six-manifold has two conformal Killing spinors $\eps_I$, but now we wlll in addition assume that there is a spacelike Killing vector field $v^M$ such that 
\bea
\L_v \eps_I &=& 0
\eea
Dimensional reduction is implemented by imposing that the Lie derivatives $\L_v$ vanish on all the fields, but in addition we will make the gauge choice
\bea
A_M v^M &=& 0
\eea
A discussion about how this can be generalized to allow for other gauge choices can be found in \cite{Gustavsson:2023kqi}. We will assume that $g^2 = v^M v_M > 0$ everywhere. This enables us to define another vector field $u_M = v_M / g^2$. Under these assumptions, the supersymmetry variations of the dimensionally reduced theory are given by 
\bea
\delta \phi &=& - i \bar\eps^I \lambda_I\cr
\delta A_N &=& i \bar\eps^I \Gamma_{MN} \lambda_I v^M\cr
\delta \lambda_I &=& - \frac{1}{2} \Gamma^{MNP} \eps_I F_{MN} u_P + \Gamma^M \eps_I D_M \phi + 4 \eta_I \phi + \frac{i}{2} \Gamma_M (\sigma^{ij})_I{}^J \eps_J [\phi^i,\phi^j] v^M 
\eea
and 
\bea
\delta \phi^i &=& i \bar\eps^I (\sigma^i)_{IA} \psi^A\cr
\delta \psi^A &=& - \Gamma^M (\sigma^i)^{AI} \eps_I D_M \phi^i - 4 (\sigma^i)^{AI}\eta_I \phi^i - i \Gamma_M (\sigma^i)^{AI} \eps_I [\phi^i,\phi] v^M
\eea
or, if we define,
\bea
q^{AI} &=& \phi^i (\sigma^i)^{AI}\cr
\phi^i &=& \frac{1}{2} (\sigma^i)_{IA} q^{AI}
\eea
then they become %are
\bea
\delta \phi &=& - i \bar\eps^I \lambda_I\cr
\delta A_N &=& i \bar\eps^I \Gamma_{MN} \lambda_I v^M\cr
\delta \lambda_I &=& - \frac{1}{2} \Gamma^{MNP} \eps_I F_{MN} u_P + \Gamma^M \eps_I D_M \phi + 4 \eta_I \phi + \frac{i}{2} \Gamma_M \eps_J [q_{IA},q^{AJ}] v^M 
\eea
and 
\bea
\delta q^{AI} &=& 2 i \bar\eps^I \psi^A\cr
\delta \psi^A &=& - \Gamma^M \eps_I D_M q^{AI} - 4 \eta_I q^{AI} - i \Gamma_M \eps_I [q^{AI},\phi] v^M
\eea

\subsection{The cohomological formulation}
We will now bring these into a cohomological form following \cite{Kallen:2012va}. We expand the tensor multiplet spinor field as
\ben
\lambda_I &=& \frac{1}{4 \N^2} \Theta^{MNP,J}{}_I U_P U^Q \Gamma_Q \eps_J \chi_{MN} + \frac{1}{\N} \Gamma^M \eps_I \Psi_M\label{invtensor}
\een
where
\bea
\chi_{MN} &=& \bar\eps^I\t\Gamma_{MN} \lambda_I\cr
\Psi_N &=& \bar\eps^I \Gamma_{MN} \lambda_I U^M
\eea
Here we define traceless gamma matrices as $\t\Gamma_M = \Gamma_M - \frac{1}{\N} \(V_M U^N + U_M V^N\) \Gamma_N$ and $\t\Gamma_{MN} = \t\Gamma_{[M} \t\Gamma_{N]}$. Then the field $\chi_{MN}$ satisfies
\bea
\chi_{MN} &=& \frac{1}{2\N} \eps_{MN}{}^{PRST} V_P U_R \chi_{ST}\cr
\chi_{MN} U^M &=& 0\cr
\chi_{MN} V^M &=& 0
\eea
and, thus, it has $3$ components, while $\Psi_M$ that satisfies
\bea
\Psi_M U^M &=& 0
\eea
has $5$ components. In total we have $3+5 = 8$ components, which agrees with the number of spinor components in the two spinors $\lambda_I$ for $I = 1,2$. 

For convenience below, we will also introduce 
\bea
\Psi_{MN} &=& \bar\eps^I \Gamma_{MN} \lambda_I\cr
\Psi &=& \frac{1}{\N} \Psi_M V^M
\eea

For the hypermultiplet, we first define two new spinor fields as
\bea
\Psi^A &=& V^M \Gamma_M \psi^A\cr
\Up^A &=& U^M \Gamma_M \psi^A
\eea
which can be inverted as
\bea
\psi^A &=& \frac{1}{2\N} U^M \Gamma_M \Psi^A + \frac{1}{2\N} V^M \Gamma_M \Up^A\label{invhyper1}
\eea
and then we use these to define fermionic fields in the cohomological formulation as 
\bea
\Psi^{AI} &=& \frac{1}{\N} \bar\eps^I \Gamma_M \Psi^A U^M\cr
\Up^{AI}_M &=& \bar\eps^I \Gamma_M \Up^A
\eea

For the vector multiplet, we get
\bea
\delta \phi &=& - i \Psi\cr
\delta A_N &=& i \chi_{MN} v^M\cr
\delta \Psi_{MN} &=& V^P u_P F_{MN} + V^P F_{PM} u_N - V^P F_{PN} u_M \cr
&& - \frac{1}{2} V_P \eps_{MN}{}^{PRST} F_{RS} u_T\cr
&& - \partial_M \(V_N \phi\) + \partial_N \(V_M \phi\)\cr
&& +  \frac{i}{2} \Theta_{MNP}{}^I{}_J [q_{IA},q^{AJ}] v^P\cr
\delta \Psi_N &=& 3 V^P U^M \(F_{[MN} u_{P]} - \frac{1}{6} \eps_{MNP}{}^{RST} F_{RS} u_T\)\cr
&& - U^M \partial_M (V_N \phi) + U^M \partial_N \(V_M \phi\)\cr
&& + \frac{i}{2} \Theta_{MNP}{}^I{}_J [q_{IA},q^{AJ}] v^P U^M\cr
\delta \Psi &=& \L_V \phi + \frac{1}{3} \nabla_M V^M \phi
\eea
and, for the hypermultiplet, we get
\bea
\delta q^{AI} &=& i \Psi^{AI}\cr
\delta \Psi^{AI} &=& - \(\L_V q^{AI} + \frac{1}{3} \nabla_M V^M q^{AI} + 8 R^I{}_J q^{AJ}\) + i [V^M A_M + v^M V_M \phi,q^{AI}]\cr
\delta \Up_Q^{AI} &=& - \frac{1}{2} V_Q U^N D_N q^{AI}\cr
&& - \frac{1}{2} U_Q V^N D_N q^{AI}\cr
&& + \frac{\N}{2} D_Q q^{AI}\cr
&& - \Theta_{QNP}{}^I{}_J U^N D^P q^{AJ}\cr
&& - 4 U^N \bar\eps^I \Gamma_Q \Gamma_N \eta_J q^{AJ}\cr
&& - i [q^{AJ},\phi] \(\Theta_{QNP}{}^I{}_J v^P U^N + \frac{1}{2} \(V^P v_P U_Q + V_Q v^P U_P - v_Q \N\)\)  
\eea
When we formulate the supersymmetry variations in this cohomological form, the supersymmetry parameter $\eps$ has been absorbed in the fermionic fields. That means that, as we perform a supersymmetry variation of such a fermionic field, we shall find a bilinear in the supersymmetry parameter. These bilinears that appear in the variations above are $V_M = \bar\eps^I \Gamma_M \eps_I$ and $\Theta_{MNP}{}^I{}_J = \bar\eps^I \Gamma_{MNP} \eps_J$ and we also have additional bilinears by using $\eta_I$. Those are $\bar\eps^I \eta_J = \frac{1}{24} \delta^I_J \nabla_M V^M + R^I{}_J$ where $R^I{}_I = 0$ and $\bar\eps^I \Gamma_{MN} \eta_J = - \frac{1}{4} \delta^I_J \(\nabla_M V_N - \nabla_N V_M\) + \t{\bar\eps^I \Gamma_{MN} \eta_J}$ where tilde is used to indicate the traceless part, which means that its contraction with $\delta^J_I$ vanishes. %is zero. 

We note that we assume the existence of a Killing vector $v^M$ and we assume that there are at least two conformal Killing spinors that satisfy $\L_v \eps_I = 0$. These assumptions restrict the class of geometries of Lorentzian six-manifolds.

For a generic curved six-manifold that admits say $n$ complex conformal Killing spinors, it is in general a difficult problem to count how many, if any, of those solutions will also satisfy $\L_v \eps_I = 0$. We may restrict ourselves to flat $\mb{R}^{1,4} \times S^1$ for the dimensional reduction insofar as the nonabelian structure is concerned. Then we have $8$ Poincare supercharges, all of which survive under the dimensional reduction. 

However, other interesting features will be seen by allowing for somewhat more general circle bundle geometries than just $\mb{R}^{1,4} \times S^1$. In particular we will be able to see that a 6d loop space formulation can exist in a consistent manner by showing that a certain commutator term that is problematic to uplift to loop space, is actually zero exactly because of a geometric property. Namely equation (\ref{geometr}) below, holds for generic circle bundles on which we perform the dimensional reduction.

\section{The six-dimensional $(1,0)$ tensor multiplet}\label{section6}
Also for the abelian six-dimensional theories, there is a corresponding cohomological formulation whose nonabelian generalization is not known. We will use the abelian cohomological formulation as a stepping stone to obtain the corresponding nonabelian supersymmetry variations in loop space. Towards this goal, we will proceed as follows.

In section \ref{One} we obtain the abelian supersymmetry variations in the cohomological form. 

In section \ref{Two} we obtain the abelian supersymmetry variations in loop space. 

In section \ref{miniloop} we obtain the abelian supersymmetry variations in the mini loop space. Here we make the nontrivial observation that a Lie derivative that we compute in the mini loop space agrees with the corresponding Lie derivative that we compute in the loop space when we restrict the loop to wrap the orbit of the Killing vector field $v$. 

In section \ref{mini-non-tensor} we obtain the nonabelian supersymmetry variations in the mini loop space directly from five-dimensional super Yang-Mills. 

In section \ref{Three} we finally conjecture the nonabelian generalization of the supersymmetry variations in loop space. These are such that they reduces correctly to the nonabelian supersymmetry variations in the mini loop space when we restrict the loops to wrap the orbits of the Killing vector $v$. We also check that these supersymmetry variations close among themselves up to a gauge variation with the gauge parameter 
\bea
\Lambda &=& - i \B
\eea
where $\B$ is a certain nonabelian loop space field. While we do not have an explicit realization of our nonabelian loop space fields in terms of some nonabelian spacetime fields, we have a corresponding definition for the abelian gauge group, where 
\bea
\B &=& \int ds \(B_{NM} V^N - V_M \phi\) \dot{C}^M
\eea
is obtained from integrating the two-form gauge potential $B_{MN}$ and the scalar field $\phi$ around the loop in a reparametrization invariant way by making use of both the lightlike Dirac current $V_M$ and the tangent vector $\dot{C}^M$ of the loop.

\subsection{The cohomological formulation}\label{One}
Here we reformulate the abelian tensor multiplet in its cohomological form. We find that it is convenient but not strictly necessary for the cohomological formulation, to introduce the following supersymmetry singlet field, 
\bea
B_N &=& B_{MN} V^M - V_N \phi
\eea
We also introduce the corresponding dual one-form potential
\bea
B^{\vee}_N &=& B_{MN} U^M
\eea
that however does not involve the scalar field. We will use the notation $\t{B}_{MN}$ to denote the traceless part\footnote{It should be noted that we use two different notions of a 'trace' in our paper. For further clarification of these traces we refer to the introduction paragraph of the appendix.} of $B_{MN}$, which means that $\t{B}_{MN} V^M = 0$ and $\t{B}_{MN} U^M = 0$. We notice that $B^{\vee}_M U^M = 0$ and $B_M V^M = 0$. We next introduce the fermionic tensors and scalar fields that are essential for the cohomological reformulation, 
\bea
\Psi_{MN} &=& \bar\eps^I \Gamma_{MN} \lambda_I\cr
\Psi_N &=& \bar\eps^I \Gamma_{MN} \lambda_I U^M\cr
\Psi &=& \frac{1}{\N} \bar\eps^I \Gamma_{MN} \lambda_I U^M V^N
\eea
These definitions are analogous to the definitions we made above in 5d super Yang-Mills. The supersymmetry variations for the abelian tensor multiplet in terms of these fields become
\ben
\delta \phi &=& - i \Psi\cr
\delta B_{MN} &=& i \Psi_{MN}\cr
\delta B_N &=& 0\cr
\delta B^{\vee}_N &=& i \Psi_N\cr
\delta \Psi_{MN} &=& - \nabla_M \(V_N \phi\) + \nabla_N \(V_M \phi\)\cr
&& - V^P H_{PMN}\cr
&& + \frac{1}{2} V^P H_{PMN}^-\cr
\delta \Psi_N &=& - V^P U^M H_{PMN} + \frac{1}{2} V^P U^M H_{PMN}^-\cr
&& - V_N U^M \nabla_M \phi + \N \nabla_N \phi + U^M \(\nabla_N V_M - \nabla_M V_N\) \phi\cr
\delta \Psi &=& \L_V \phi + \frac{1}{3} \nabla_M V^M \phi\label{precise}
\een
It shall be noted that the supersymmetry parameters $\eps_I$ do not appear in the supersymmetry variations. So to get back to the original supersymmetry variations where $\eps_I$ appear, we need to provide the parameters $\eps_I$ in addition to the supersymmetry variations above.

\subsection{The loop space formulation}\label{Two}
We define loop space as the space of loops in the following sense. A loop is a defined as a smooth map from $S^1$ into spacetime, $s \mapsto C^M(s)$. Different parametrizations of the loop correspond to one and the same loop $C$. Since we do not fix a base point, our loop space is a free loop space. A reparametrization invariant metric on free loop space can be introduced to make it a metric space \cite{U}, although for our purposes of writing down supersymmetry variations we do not need a metric. Fields on loop space depend on the loop $C$ but not on the way that the loop $C$ is parametrized. In other words, the fields on loop space (or the loop space fields) shall be reparametrization invariant. Since we use $\delta$ to denote a supersymmety variation, we will use the notation $\delta_C$ with a subscript $C$ for the differential operator that acts on a loop space field. However, we will use the short notation $\delta C^M(s)$ instead of $\delta_C C^M(s)$ to denote a one-form differential in loop space. 

We define loop space fields as
\bea
\A &=& \int ds \t{B}_{MN} \dot{C}^M \delta C^N\cr
\B &=& \int ds B_M \dot{C}^M\cr
\C &=& \int ds B^{\vee}_M \dot{C}^M\cr
\Phi_0 &=& \int ds \phi V_M \dot{C}^M\cr
\Psi_0 &=& \int ds \Psi V_M \dot{C}^M\cr
\Psi_1 &=& \int ds \chi_{MN} \dot{C}^M \delta C^N\cr
\Up_0 &=& \int ds \Psi_M \dot{C}^M
\eea
One may easily check that they are all reparametrization invariant. We will assume that $H_{MNP}$ is selfdual, so that $H^-_{MNP} = 0$. With this assumption, we then obtain the following supersymmetry variations in loop space
\bea
\delta \A &=& i \Psi_1\cr
\delta \B &=& 0\cr
\delta \C &=& i \Up_0\cr
\delta \Psi_1 &=& - \L_V \A - \delta_C \B\cr
\delta \Phi_0 &=& - i \Psi_0\cr
\delta \Psi_0 &=& \L_V \Phi_0\cr
\delta \Up_0 &=& \L_U \B - \L_V \C
\eea
where, to get the form of supersymmetry variation for $\Up_0$ as stated, %holds only if 
we used the fact %assume 
that $U^M$ and $V^M$ are commuting vector fields in the sense of having a vanishing Lie bracket,
\bea
U^M \nabla_M V^N - V^M \nabla_M U^N &=& 0
\eea
The on-shell closure relations on all these fields are
\bea
\delta^2 &=& - i \L_V + \delta_{-i \B}
\eea
where the second term is a gauge variation with gauge parameter $-i \B$. A general infinitesimal gauge transformation with gauge parameter 
\bea
\Lambda &=& \int ds \Lambda_M \dot{C}^M
\eea
is given by 
\bea
\delta A &=& \delta_C \Lambda\cr
\delta \B &=& - \L_V \Lambda\cr
\delta \C &=& - \L_U \Lambda
\eea
while it acts trivially on the matter fields in the abelian theory.

We remark that closure on the pair $\Up_0$ and $\C$ is automatic without using any equation of motion,
\bea
\delta^2 \C &=& - i \L_V \C + \L_U (i \B)\cr
\delta^2 \Up_0 &=& - i \L_V \Up_0
\eea
We also have 
\bea
\delta^2 \B &=& 0\cr
&=& - i \L_V \B - \L_V (-i \B) 
\eea
Since off-shell closure is also true for all the other fields, this means that in loop space we have off-shell supersymmetry. This off-shell closure can be understood as an effect of viewing $H^-_{MNP} = 0$ as a constraint rather than incorporating this as a subtle equation of motion in some way in the loop space. How to possibly deal with selfduality in loop space is a rather deep problem that we will not discuss here.

\subsection{Mini loop space}\label{miniloop}
Dimensional reduction of the six-dimensional theory results in a five-dimensional super Yang-Mills theory that we can express in terms of local spacetime fields. We can get local spacetime fields from loop space fields in six dimensions if we take the loops to wrap the compact circle along which we dimensionally reduce the theory. In our formulation of the dimensionally reduced theory, we introduced a spacelike Killing vector field $v^M$. We now define the corresponding loops such that  
\bea
\frac{dC^M}{ds} &=& v^M(C(s))
\eea
The loop space fields now become local spacetime fields in five dimensions that are given by
\ben
\A &=& A_M dx^M\cr
\B &=& - v^M V_M \phi - A_M V^M\cr
\C &=& - A_M U^M\cr
\Phi_0 &=& \phi V_M v^M\cr
\Psi_0 &=& \Psi V_M v^M\cr
\Psi_1 &=& \chi_{MN} v^M dx^N\cr
\Up_0 &=& \Psi_M v^M\label{vSYMtomini}
\een
Their supersymmetry variations are 
\bea
\delta \A &=& i \Psi_1\cr
\delta \B &=& 0\cr
\delta \C &=& i \Up_0\cr
\delta \Phi_0 &=& - i \Psi_0\cr
\delta \Psi_0 &=& \L_V \Phi_0\cr
\delta \Psi_1 &=& - \L_V \A - d \B\cr
\delta \Up_0 &=& \L_U \B - \L_V \C
\eea
To show these variations in this restricted mini loop space, we perform some computations that look quite different from corresponding computations in the 6d loop space and yet the resulting supersymmetry variations look the same. This is not a surprise since by restricting to a subset of loops that wrap around a spatial circle we do not expect to get a different result from what we get in the full loop space. 

To show the supersymmetry variation of $\Psi_0$ we compute the Lie derivative in the mini loop space,
\ben
\L_V \Phi_0 &=& \L_V \(\phi V_M v^M\)\cr
&=& V^N \nabla_N \(\phi V_M v^M\)\cr
&=& \L_V \phi V_M v^M + \phi V^N \nabla_N \(V_M v^M\)\cr
&=& \L_V \phi V_M v^M + \phi \(V^N \nabla_N V_M v^M + \frac{1}{2} V^N V^M \(\nabla_N v_M + \nabla_M v_N\)\)\cr
&=& \(\L_V \phi + \frac{1}{3} \nabla_P V^P \phi\) V_M v^M\label{T1}
\een
To go from the fourth to the fifth line we use the Killing vector equation for $v_M$,
\bea
\nabla_M v_N + \nabla_N v_M &=& 0
\eea
as well as the conformal Killing vector equation for $V_M$ and the fact that $V^N$ is lightlike. Let us now contrast this result with the corresponding result in the full loop space where we get
\bea
\L_V \Phi_0 &=& \int ds \(\L_V \Phi + \frac{1}{3} \nabla_P V^P \phi\) V_M \dot{C}^M 
\eea
Of course we do not use any Killing vector equation for the tangent vector of the loop in the full loop space. So clearly the two computations are very different from each other, and yet the final results are essentially the same.

To show the variation of $\Up_0$ we first note that
\bea
\delta \Up_0 &=& V^P U^M F_{PM} - \L_U\(V_N v^N \phi\) + U^M \L_v\(V_M \phi\)
\eea
where the last term shall be put to zero by the dimensional reduction. The remaining terms  we can express in terms of $\B$ and $\C$ as
\bea
\delta \Up_0 &=& \L_U \B - \L_V \C
\eea
To see how this work, let us just expand the right-hand side,
\bea
\L_U \B - \L_V \C &=& U^M \nabla_M \(- V^N A_N - ...\) - V^M \nabla_M \(- U^N A_N\)\cr
&=& - U^M V^N F_{MN} + ... - \(U^M \nabla_M V^N - V^M \nabla_M U^N\) A_N
\eea 
We thus get the desired result using the fact %if we assume 
that the Lie bracket 
\bea
[U,V]^N &=& U^M \nabla_M V^N - V^M \nabla_M U^N 
\eea
is zero.

The show the variation of $\Psi_{MN}$ we note that
\bea
\delta \Psi_{MN} v^M &=& - \L_V A_N - \nabla_N \B - \L_v\(V_N \phi\)
\eea 
and, then by dimensional reduction, the last term is put to zero and the result can be recast in the above form.

\subsection{Nonabelian generalization in mini loop space}\label{mini-non-tensor}
For the vector multiplet we define $\chi_{MN} = \bar\eps^I \t\Gamma_{MN} \lambda_I$ for which there is a commutator term in the nonabelian generalization
\ben
\delta \chi_{MN} &=& ... + \frac{i}{2} \Theta_{MNP}{}^I{}_J [q_{IA},q^{AJ}] v^P\label{killed1}
\een
We then get 
\ben
\delta \chi_{MN} v^N &=& ... + 0\label{killed2}
\een
which implies that in none of the mini loop space fields $\Psi_1 = \chi_{MN} v^M dx^N$, $\Up_0 = \chi_{MN} v^M V^N$ or $\Psi_0 = \Psi V_M v^M$ are we able to see this commutator. The full result in the nonabelian case becomes
\bea
\delta \chi_{MN} v^M &=& - \L_V A_N - \L_v(V_N \phi) + D_N (V^M A_M + v^M V_M \phi)
\eea
where, by the dimensional reduction, we shall put $\L_v(V_N \phi) = 0$. This results in the following nonabelian generalization for the supersymmetry variation of the corresponding mini loop space fermion field $\Psi_1$, 
\bea
\delta \Psi_1 &=& - \L_V \A - D \B\cr
&=& - \L_V \A - d \B + i [\A,\B]
\eea
Similarly, for supersymmetry variation of the mini loop space fermion field $\Up_0$, we obtain its nonabelian generalization as
\bea
\delta \Up_0 &=& U^M \L_V A_M + U^M D_M \B\cr
&=& - \L_V \C + \L_U \B - i U^M [A_M,\B]\cr
&=& - \L_V \C + \L_U \B + i [\C,\B]
\eea

\subsection{The nonabelian tensor multiplet in loop space}\label{Three}
We would like to uplift the supersymmetry variations in the mini loop space to the full loop space in order to describe the six-dimensional nonabelian tensor multiplet. We propose the following supersymmetry variations in the loop space,
\begin{align}
\delta \A &= i \Psi_1\cr
\delta \B &= 0\cr
\delta \C &= i \Up_0\cr
\delta \Phi_0 &= - i \Psi_0\cr
\delta \Psi_1 &= - \L_V \A - \delta_C \B + i [\A,\B]\cr
\delta \Psi_0 &= \L_V \Phi_0 + i [\B,\Phi_0]\cr
\delta \Up_0 &= \L_U \B - \L_V \C - i [\B,\C]
\end{align}
We can not derive these variations from the nonabelian tensor multiplet in spacetime since these are not known. But we can perform some consistency checks of our proposed supersymmetry variations. Firstly, they reduce to the abelian supersymmetry variations for an abelian gauge group. Secondly, they reduce to the nonabelian supersymmetry variation in the mini loop space. Thirdly, the supersymmetry variations close among themselves such that  
\bea
\delta^2 &=& - i \L_V + \delta_{\Lambda}
\eea
when we act twice on anyone of the loop space fields. Here the second term is a gauge transformation with gauge parameter 
\bea
\Lambda &=& - i \B
\eea
A general infinitesimal gauge transformation acts on the fields as
\bea
\delta \A &=& D_C \Lambda\cr
\delta \B &=& - \L_V \Lambda - i [\B,\Lambda]\cr
\delta \C &=& - \L_U \Lambda - i [\C,\Lambda]\cr
\delta \Phi &=& - i [\Phi,\Lambda]
\eea
where $\Phi$ represent any matter field. Here we have introduced a gauge covariant derivative 
\bea 
D_C \Phi &=& \delta_C \Phi - i [\A,\Phi]
\eea
on loop space. Interestingly, this form of the loop space covariant derivative has appeared in the literature previously in \cite{Baez:2004in}.

\section{The six-dimensional $(1,0)$ hypermultiplet}\label{section7}
We proceed with the same steps with the hypermultiplet as for the tensor multiplet. In section \ref{OneH} we obtain the abelian supersymmetry variations in the cohomological form. This reformulation is more complicated for the hypermultiplet than for the tensor multiplet. But as a guide we have the reference \cite{Kallen:2012va}. There the corresponding cohomological formulation was obtained for the hypermultiplet in five dimensions. In section \ref{TwoH} we use the cohomological formulation to obtain the abelian supersymmetry variations in loop space. In section \ref{mini-non-hyper} we present the fields in the mini loop space. In section \ref{mini-non-hyperH} we obtain the nonabelian supersymmetry variations in the mini loop space from five-dimensional super Yang-Mills. In section \ref{ThreeH} we finally conjecture the nonabelian generalization of the supersymmetry variations in loop space, which are such that they reduce correctly to the nonabelian supersymmetry variations in the mini loop space when we restrict the loops to wrap the orbits of the Killing vector $v$ and they close among themselves up to a gauge variation with the same gauge parameter 
\bea
\Lambda &=& - i \B
\eea
as for the tensor multiplet. In the nonabelian generalization, the hypermultiplet couples to the tensor multiplet in a nontrivial way and the gauge parameter has to agree for consistency.

\subsection{The cohomological formulation}\label{OneH}
In this section we obtain the cohomological formulation of the supersymmetry variations for the abelian six-dimensional hypermultiplet. We start with defining fermionic scalar fields as 
\bea
\Psi^{AI} &=& 2 \N \bar\eps^I \psi^A
\eea
that are labeled by $A = 1,...,N_f$ where $N_f$ is the number of hypermultiplets, and $I = 1,2$ is the $SU(2)$ R symmetry index. This map from the four-component hypermultiplet spinors $\psi^A$ to the two-component scalar fields $\Psi^{AI}$ labeled by $I = 1,2$ for each hyper labeled by $A = 1,...,N_f$, can not be invertible since we map four components into two. So to get an invertible map, we need to add two more fermionic fields. There seems to be no covariant way of adding two more fermionic fields. So instead, we will add a fermionic vector field 
\bea
\Up_M^{AI} &=& \bar\eps^I \Gamma_M \Gamma_N \psi^A U^N
\eea
Then all these components $\Psi^{AI}$ and $\Up_M^{AI}$ can not be independent, but the exact relation between these components seems difficult to express in a closed form. However, we can immediately reduce the number of components in $\Up_M^{AI}$ from six down to four by noting that $V^M \Up_M^{AI} = 0$ and $U^M \Up_M^{AI} = 0$.

The story for the fields $q^{AI}$ and $\Psi^{AI}$ is very neat. We have the following supersymmetry variations 
\bea
\delta q^{AI} &=& i \Psi^{AI}\cr
\delta \Psi^{AI} &=& - \(\L_V q^{AI} + \frac{1}{3} \nabla_M V^M q^{AI} + 8 R^I{}_J q^{AJ}\)
\eea
We have the closure relations
\bea
\delta^2 q^{AI} &=& - i \L_V q^{AI} - \frac{1}{3} \nabla_M V^M q^{AI} - 8 i R^I{}_J q^{AJ}\cr
\delta^2 \Psi^{AI} &=& - i \L_V \Psi^{AI} - \frac{1}{3} \nabla_M V^M \Psi^{AI} - 8 i R^I{}_J \Psi^{AJ}
\eea
for which we do not use any equation of motion. We uplift this to loop space by first defining the loop space fields
\bea
\Phi_0^{AI} &=& \int ds \dot{C}^M V_M q^{AI}\cr
\Psi_0^{AI} &=& \int ds \dot{C}^M V_M \Psi^{AI}
\eea
Then the supersymmetry variations are
\bea
\delta \Phi_0^{AI} &=& i \Psi_0^{AI}\cr
\delta \Psi_0^{AI} &=& - \L_V \Phi_0^{AI} - 8 R^I{}_J \Phi_0^{AJ}
\eea
The closure relations are
\bea
\delta^2 \Phi_0^{AI} &=& - i \L_V \Phi_0^{AI} - 8 i R^I{}_J \Phi_0^{AJ}\cr
\delta^2 \Psi_0^{AI} &=& - i \L_V \Psi_0^{AI} - 8 i R^I{}_J \Psi_0^{AJ}
\eea
In appendix \ref{RIJ} we show that 
\bea
R^I{}_J &=& \bar\eps^I \eta_J - \frac{1}{2} \delta^I_J \bar\eps^K \eta_K
\eea
is covariantly constant and can therefore be taken outside the integral over the loop. In appendix \ref{LieDer} we show that 
\bea
\L_V \Phi_0^{AI} &=& \int ds \dot{C}^M V_M \(V^N \nabla_N q^{AI} + \frac{1}{3} \nabla_N V^N q^{AI}\)
\eea
This completes the story for these fields in the hypermultiplet, for the abelian case. 

Now we will address the question of what additional cohomological fields we should include in this story. To this end, we follow %closely 
\cite{Kallen:2012va} closely. The first step is to convert all fields into spinor fields. We thus define a bosonic spinor field
\bea
Q^A &=& q^{AI} \eps_I
\eea
For the fermions, we define new fermionic fields as
\bea
\Psi^A &=& V^M \Gamma_M \psi^A\cr
\Up^A &=& U^M \Gamma_M \psi^A
\eea
The supersymmetry variations for these spinorial bosonic and fermionic fields are
\bea
\delta Q^A &=& \frac{i}{2} \Psi^A\cr
\delta \Psi^A &=& - 2 \L_V Q^A - \frac{3}{2} \nabla_M V^M Q^A\cr
\delta \Up^A &=& - 2 U^M  \Gamma_M \Gamma^N \nabla_N Q^A - \frac{2}{\N} \Gamma^{MN} Q^A U_M U^P \nabla_P V_N
\eea
We have been unable to recast the supersymmetry variation of $\Up^A$ in a nice from, but nevertheless the closure relations are of a nice form, 
\bea
\delta^2 Q^A &=& - i \L_V Q^A - \frac{i}{4} \nabla_M V^M Q^A\cr
\delta^2 \Psi^A &=& - i \L_V \Psi^A - \frac{i}{4} \nabla_M V^M \psi^A\cr
\delta^2 \Up^A &=& - i \L_V \Up^A - \frac{i}{4} \nabla_M V^M \Up^A\cr
&& + \frac{i}{2} \(\Gamma^M V_M \Gamma^N \nabla_N \Up^A - 2 U^M \nabla_M \Psi^A - \Gamma^M \Gamma^N \nabla_M U_N \Psi^A + 2 V^N \nabla_M U_N \Gamma^M \psi^A\)\cr
&& + i \N \Gamma^M \nabla_M \psi^A
\eea
Now we see that to have closure on $\Up^A$\negthinspace, we need to use the equation of motion for the fermionic fields. Since the R-symmetry index $I$ in the definition of $Q^A$ is fully contracted, we see that there does not appear any R-symmetry rotations in these closure relations. 

The next step, following  \cite{Kallen:2012va} closely, is to convert these spinorial fields into $p$-form fields. This step is very helpful for us because we understand how to integrate a $p$-form over a loop to give us a $p-1$ form in loop space, but we do not know how to integrate a fermionic field over a loop. 

We define one-form and three-form fields for the scalars
\bea
Q^{AI}_M &=& \bar\eps^I \Gamma_M Q^A\cr
Q^{AI}_{MNP} &=& \bar\eps^I \Gamma_{MNP} Q^A
\eea
and, similarly for the fermions,
\bea
\Psi^{AI}_M &=& \bar\eps^I \Gamma_M \Psi^A\cr
\Psi^{AI}_{MNP} &=& \bar\eps^I \Gamma_{MNP} \Psi^A\cr
\Up^{AI}_M &=& \bar\eps^I \Gamma_M \Up^A\cr
\Up^{AI}_{MNP} &=& \bar\eps^I \Gamma_{MNP} \Up^A
\eea
We have
\bea
Q^{AI}_M &=& \bar\eps^I \Gamma_M Q^A\cr
&=& \bar\eps^I \Gamma_M \eps_J q^{AJ}\cr
&=& \frac{1}{2} V_M q^{AI}
\eea
This relation can be inverted,
\bea
q^{AI} &=& \frac{2}{\N} Q_M^{AI} U^M\cr
&=& \frac{2}{\N} Q^{AI}
\eea
We also notice that $Q_{MNP}^{AI}$ is not an independent field, but can be obtained from the one-form as
\bea
Q_{MNP}{}^I{}_J &=& \frac{2}{\N} \Theta_{MNP}{}^I{}_J U^Q Q_Q^{AJ}
\eea
A similar relation holds for the fermionic field $\Psi^{AI}_M$ but the computation that is required to see this is somewhat different,
\bea
\Psi^{AI}_M &=& \bar\eps^I \Gamma_M \Psi^A\cr
&=& \bar\eps^I \Gamma_M \Gamma_N \psi^A V^N\cr
&=& 2 \bar\eps^I \psi^A V_M
\eea
Here in the last step we have used $\Gamma_N V^N \eps = 0$. Now we also get
\bea
\Psi^{AI} &=& 2 \N \bar\eps^I \psi^A
\eea
Let us also notice that, if we assume that $\Psi^A = \frac{2}{\N} \eps_J \Psi^{AJ}$\negthinspace\negthinspace, then we immediately get
\bea
\Psi^{AI}_M &=& \frac{1}{\N} V_M \Psi^{AI}
\eea
which is consistent with the above. So we recover the Weyl spinor as
\ben
\Psi^A &=& \frac{2}{\N} \eps_I \Psi^{AI}\label{invhyper2}
\een
where we define 
\bea
\Psi^{AI} &=& \Psi^{AI}_M U^M
\eea
We see that we did not need to introduce the three-form $\Psi^{AI}_{MNP}$ %and if we do, then it will
as it is not be an independent field. Likewise we did not need to introduce the full one-form $\Psi^{AI}_M$ as all information about the field sits in $\Psi^{AI}$. 

Next we notice that 
\bea
\Up^{AI}_M V^M &=& 0\cr
\Up^{AI}_M U^M &=& 0
\eea
the first relation follows from $V^M \Gamma_M \eps_I = 0$ and the second relation follows from $U^M \Gamma_M \Up^A = 0$. Nonetheless, we can again invert the relation, and get back the spinor field from just the one-form
\ben
\Up^A &=& \frac{1}{8\N} \Gamma^{MN} \eps_I \Up_M^{AI} U_N\label{invhyper3}
\een
We notice that $P_+ \Up^A = 0$ follows automatically. To see that, one may carry out the following computation
\bea
P_+ \Gamma^{MN} \eps_I \Up_M U_N &=& \{P_+,\Gamma^{MN}\} \eps_I \Up_M U_N - \Gamma^{MN} \eps_I \Up_M U_N\cr
&=& \frac{V_P U_Q}{2\N} \{\Gamma^{PQ},\Gamma^{MN}\} \eps_I \Up_M U_N 
\eea
We do not need to introduce the three-form $\Up^{AI}_{MNP}$ since this will not be an independent field. 

Let us summarize our result. We have found the following independent $p$-form fields for the hypermultiplet. There are $2N_f$ real bosonic fields
\bea
Q^{AI} &=& Q_M^{AI} U^M
\eea
for $A = 1,...,N_f$. There are $2N_f$ real fermions
\bea
\Psi^{AI} &=& \Psi^{AI}_M U^M
\eea
and then there are additional real fermions
\bea
\Up^{AI}_M
\eea
that are constrained by $\Up^{AI}_M U^M = 0$ and $\Up^{AI}_M V^M = 0$, but since these should comprise the missing $2N_f$ fermions, that means that there should be some additional hidden algebraic relations between the remaining $4 \times 2 N_f$ components $\Up^{AI}_M$ that reduce these down to $2N_f$ independent components. However, it seems difficult to express such algebraic relations in a covariant form so we will keep this redundant form for these fields.

Then the abelian hypermultiplet supersymmetry variations become 
\bea
\delta q^{AI} &=& i \Psi^{AI}\cr
\delta \Psi^{AI} &=& - \(\L_V q^{AI} + \frac{1}{3} \nabla_M V^M q^{AI} + 8 R^I{}_J q^{AJ}\)\cr
\delta \Up_Q^{AI} &=& - \frac{1}{2} V_Q \(U^N \nabla_N q^{AI} + \frac{1}{3} \nabla_N U^N q^{AI}\)\cr
&& - \frac{1}{2} U_Q \(V^N \nabla_N q^{AI} + \frac{1}{3} \nabla_N V^N q^{AI}\)\cr
&& + \frac{1}{2} \nabla_Q \(\N q^{AI}\)\cr
&& + \Theta_{QMN}{}^I{}_J \nabla^M\(U^N q^{AJ}\)\cr
\delta Q^{AI}_{MN} &=& i \Psi^{AI}_{MN}\cr
\delta \Psi^{AI}_{MN} &=& - \Theta_{MN}{}^I{}_J \L_V q^{AJ} - \frac{1}{3} \nabla_P V^P Q_{MN}^{AI}\cr
&& - 8 \Theta_{MN}{}^I{}_J R^J{}_K q^{AK} 
\eea
Here we have
\bea
\Psi^{AI}_{MN} &=& \Theta_{MN}{}^I{}_J \Psi^{AJ}\cr
Q^{AI}_{MN} &=& \Theta_{MN}{}^I{}_J q^{AJ}
\eea
where $\Theta_{MN}{}^I{}_J $ is defined in (\ref{F2}).
For how to obtain the supersymmetry variation of $\Up_Q^{AI}$ we refer to appendix \ref{UpQAI}.

\subsection{The loop space formulation}\label{TwoH}
We define loop space fields
\bea
\Phi^{AI}_0 &=& \int ds \dot{C}^M V_M q^{AI}\cr
\Phi^{\vee AI}_0 &=& \int ds \dot{C}^M U_M q^{AI}\cr
\Psi^{AI}_0 &=& \int ds \dot{C}^M V_M \Psi^{AI}\cr
\Psi^{\vee AI}_0 &=& \int ds \dot{C}^M U_M \Psi^{AI}\cr
\Up^{AI}_0 &=& \int ds \dot{C}^M \Up^{AI}_M\cr
Q^{AI}_1 &=& \int ds \dot{C}^M \delta C^N Q_{MN}^{AI}\cr
\Psi^{AI}_1 &=& \int ds \dot{C}^M \delta C^N \Psi_{MN}^{AI}
\eea
For these fields we find the following supersymmetry variations
\bea
\delta \Phi^{AI}_0 &=& i \Psi^{AI}_0\cr
\delta \Phi^{\vee AI}_0 &=& i \Psi^{\vee AI}_0\cr
\delta \Psi^{AI}_0 &=& - \L_V \Phi^{AI}_0 - 8 R^I{}_J \Phi^{AJ}_0\cr
\delta \Up^{AI}_0 &=& - \frac{1}{2} \L_U \Phi^{AI}_0 - \frac{1}{2} \L_V \Phi^{\vee AI}_0 + \delta^{\dag} Q^{AI}_1\cr
\delta Q^{AI}_1 &=& i \Psi^{AI}_1\cr
\delta \Psi^{AI}_1 &=& - \L_V Q^{AI}_1 - 8 R^I{}_J Q^{AJ}_1
\eea
To obtain the supersymmetry variation of $\Psi_1^{AI}$\negthinspace\negthinspace, we have used the following identity 
\ben
\L_V \Theta_{MN}{}^I{}_J &=& 8 \(\Theta_{MN}{}^I{}_K R^K{}_J - R^I{}_K \Theta_{MN}{}^K{}_J\) + \frac{1}{3} \nabla_PV^P \Theta_{MN}{}^I{}_J\label{unknown}
\een
To derive the identity (\ref{unknown}), we use 
\bea
\L_V \Gamma_M &=& \frac{1}{6} \Gamma^M \nabla_Q V^Q\cr
\L_V U_P &=& - \frac{1}{3} U_P \nabla_Q V^Q\cr
\L_V \eps_I &=& \frac{1}{12} \eps_I \nabla_Q V^Q + 8 \eps_J R^J{}_I\cr
\L_V \bar\eps^I &=& \frac{1}{12} \bar\eps^I \nabla_Q V^Q - 8 R^I{}_J \bar\eps^J
\eea
Furthermore we have the identity
\bea
R^I{}_K \Theta_{MN}^{KJ} &=& R^J{}_K \Theta_{MN}^{KI}
\eea
This identity is easily shown as follows. First we expand 
\bea
\bar\eps^I \eta_K \bar\eps^K \Gamma_{MNP} \eps^J &\sim & \bar\eps^I \Gamma_{MNP} \eps^J \nabla_Q V^Q
\eea
and, then, we notice that the right-hand side is symmetric in $IJ$ and the above identity %above 
immediately follows. Next we have
\bea
R^I{}_K \Theta_{MN}{}^{KJ} = \Theta_{MN}{}^{IK} R^J{}_K = -\Theta_{MN}{}^I{}_K R^{JK} = \Theta_{MN}{}^I{}_K R^{KJ}
\eea
or, in other words,
\bea
R^I{}_K \Theta_{MN}{}^K{}_J - \Theta_{MN}{}^I{}_K R^K{}_J &=& 0
\eea
For how to obtain the supersymmetry variation of $\Up_0^{AI}$ we refer to appendix \ref{UpQAI}. For how we define $\delta^{\dag}$ in loop space we refer to appendix \ref{dagger}.

\subsection{Mini loop space}\label{mini-non-hyper}
In the mini loop space that we introduced in section \ref{miniloop}, we define the hypermultiplet loop space fields as
\ben
\Phi_0^{AI} &=& v^M V_M q^{AI}\cr
\Phi_0^{\vee AI} &=& v^M U_M q^{AI}\cr
\Psi_0^{AI} &=& v^M V_M \Psi^{AI}\cr
\Psi_0^{\vee AI} &=& v^M U_M \Psi^{AI}\cr
\Up_0^{AI} &=& v^M \Up_M^{AI}\cr
Q_1^{AI} &=& Q_{MN}^{AI} v^M dx^N\cr
\Psi_1^{AI} &=& \Psi^{AI}_{MN} v^M dx^N \label{hSYMtomini}
\een
These abelian mini loop space fields can be obtained either from the loop space fields by letting the loops wrap the orbit of the Killing vector $v$. But they can also be obtained from the 5d super Yang-Mills fields in the cohomological formulation. 

\subsection{Nonabelian generalization in mini loop space}\label{mini-non-hyperH}
If we use 5d super Yang-Mills in the cohomological formulation, we can of course immediately obtain the nonabelian supersymmetry variations in the mini loop space. We find a coupling between the hypermultiplet and the tensor multiplet through a commutator term that involves the tensor multiplet loop space field
\bea
\B &=& v^M B_M
\eea
We expand the Lie derivative of the hypermultiplet scalars in a similar way as we did in (\ref{T1}) for the tensor multiplet scalar,
\bea
\L_V \Phi_0^{AI} &=& v^M V_M \L_V q^{AI} + \frac{1}{3} \nabla_M V^M \Phi_0^{AI}
\eea
We get
\bea
\delta \Phi_0^{AI} &=& i \Psi_0^{AI}\cr
\delta \Phi_0^{\vee AI} &=& i \Psi_0^{\vee AI}\cr
\delta \Psi_0^{AI} &=& - \L_V \Phi_0^{AI} - 8 R^I{}_J \Phi_0^{AJ} - i [\B,\Phi_0^{AI}]\cr
\delta \Up_0^{AI} &=& - \frac{1}{2} V^M D_M \Phi^{AI}_0 - \frac{1}{2} U^M D_M \Phi^{\vee AI}_0 - D^N \(\Theta_{QMN}{}^I{}_J v^Q U^M q^{AJ}\)\cr
&& - i \(V_M U_N v^M v^N - \frac{1}{2} \N v_M v^M\) [q^{AI},\phi]\cr
\delta Q_{MN}^{AI} &=& i \Psi_{MN}^{AI}\cr
\delta \Psi_{MN}^{AI} &=& - \L_V Q_{MN}^{AI} - i [\B,Q_{MN}^{AI}]\cr
&& + \(\L_V \Theta_{MN}{}^I{}_K - 8 \Theta_{MN}{}^I{}_J R^J{}_K - \frac{1}{3} \nabla_P V^P \Theta_{MN}{}^I{}_J \) q^{AJ}
\eea
We would like to have
\bea
\delta \Psi_{MN}^{AI} &=& - \L_V Q_{MN}^{AI} - 8 R^I{}_J Q_{MN}^{AJ} - i [\B,Q_{MN}^{AI}]
\eea
To get this result, we use the identity (\ref{unknown}). Let us focus on the commutator term in this variation for the nonabelian case,
\bea
\delta \Up^{AI} &=& ... - i [q^{AI},\phi] \(v^M V_M v^P U_P - \frac{1}{2} g^2 \N\)  \label{commhyp}
\eea
Here it seems impossible to express this commutator in terms of loop space fields, unless we restrict ourselves to geometries where 
\ben
v^M v^N \(V_M U_N - \frac{1}{2} g_{MN} \N\) &=& 0\label{geometr}
\een
Is this another identity, or is it a restriction on the six-manifold that we need to impose in order to have an uplift to loop space? Let us consider a circle bundle metric of the form
\bea
ds^2 &=& - dt^2 + g_{ij} dx^i dx^j + R^2 (dy + \kappa_i dx^i)^2
\eea
Then $v^M = (0,0,0,0,0,1)$, $V^M = (R,0,0,0,0,1)$ and $U^M = (R,0,0,0,0,-1)$. We get $v_M = (0,R^2\kappa_i,R^2)$ and $\N = - 2 R^2$. Then 
\bea
v_M v_N \(V^M U^N - \frac{1}{2} g^{MN} \N\) &=& R^4\(V^5 U^5 - \frac{1}{2} (g^{55} + \kappa_i \kappa^i) \N\)\cr
&=& R^4 \(-1 - \frac{1}{2} \frac{1}{R^2} (-2 R^2)\)\cr
&=& 0
\eea
so (\ref{geometr}) is satisfied. Since we need a circle bundle geometry in order to have dimensional reduction, we conclude that (\ref{geometr}) is indeed an identity rather than a condition that we shall impose on the six-manifold. 

\subsection{The nonabelian hypermultiplet in loop space}\label{ThreeH}
We finally uplift the nonabelian five-dimensional mini loop space supersymmetry variations to six-dimensional loop space where we conjecture the following supersymmetry variations
\bea
\delta \Phi^{AI}_0 &=& i \Psi^{AI}_0\cr
\delta \Phi^{\vee AI}_0 &=& i \Psi^{\vee AI}_0\cr
\delta \Psi^{AI}_0 &=& - \L_V \Phi^{AI}_0 - 8 R^I{}_J \Phi^{AJ}_0 - i [\B,\Phi^{AI}_0]\cr
\delta \Up^{AI}_0 &=& - \frac{1}{2} \L_U \Phi^{AI}_0 - \frac{1}{2} \L_V \t\Phi^{AI}_0 + \delta^{\dag} Q^{AI}_1\cr
\delta Q^{AI}_1 &=& i \Psi^{AI}_1
\eea
They close on-shell on 
\bea
\delta^2 &=& - i \L_V + \delta_{\Lambda} - 8 i R^I{}_J 
\eea
where $\Lambda = - i \B$. That is, we again find the same closure relations as we had for our proposed nonabelian tensor multiplet, with the same gauge parameter.

\section{Discussion}\label{section8}
We have studied the $(1,0)$ superconformal theories in six dimensions and obtained the supersymmetry variations in loop space. We notice that there is no commutator term between two scalar fields in our supersymmetry variations in loop space. Let us look at the commutator term that appears in maximally supersymmetric Yang-Mills when we make a supersymmetry variation of the fermionic field,
\ben
\delta \psi &=& ... - \frac{1}{2} \Gamma^{AB} \Gamma_M \eps v^M [\phi^A,\phi^B]\label{M5}
\een
Here $v^M$ is a spatial Killing vector and we shall impose the constraints $\L_v \psi = 0$ and $\L_v \phi^A = 0$ for a five-dimensional theory. Now we may not have a spatial Killing vector field $v$ for a generic six-manifold that has two conformal Killing spinors $\eps_I$. Instead we have the lightlike Dirac current $V^M$ and then the only option would be to perform the dimensional reduction along $V^M$. In that dimensionally reduced theory, we do not have a commutator term of two scalar fields in the supersymmetry variation of neither the tensor multiplet fermion nor the hypermultiplet fermion \cite{Gustavsson:2024yxh}, which is in agreement with our result in loop space in this paper. The spatial Killing vector $v^M$ that we introduced in this paper was only a device to obtain the nonabelian generalization. We could have proceeded differently. We could have reduced along $V^M$ instead. We would then expect to find the exact same result again for the nonabelian generalization in the loop space. But with the latter dimensional reduction along $V^M$ it would perhaps have felt more natural that such commutator terms are absent in our supersymmetry variations in the loop space. Intuitively one may understand the absence of the commutator when reducing along $V^M$ as a consequence of the Weyl projection 
\bea
\Gamma_M \eps_I V^M &=& 0
\eea
and by replacing $v^M$ with $V^M$ and $\eps$ with $\eps_I$ in (\ref{M5}). 

Even if commutators between scalar fields are absent in the supersymmetry variations, there are commutator terms between matter fields in the Lagrangian in the five-dimensional theory we get upon dimensional reduction along $V^M$ \cite{Gustavsson:2024yxh}. It would be very interesting if such commutator terms can also be found in a corresponding loop space Lagrangian. We would thus like to find the superconformal Lagrangian in loop space. Here we encountered a problem immediately though. To construct a Lagrangian, we need a metric on the loop space, and to construct an action we need an integration measure on the loop space. The metric seems like the easier problem, the integration measure the more difficult problem. But even for the metric, this problem does not seem trivial either. What comes to mind at first is to take the metric on loop space to be given by 
\bea
g_{Ms,Ns'}(C) &=& g_{MN}(C(s)) \delta(s-t)
\eea
But this metric does not transform covariantly under reparametrizations. Perhaps one can overcome this problem by introducing a one-bein field $e(s)$ along the loop. Then one could try to define the metric on the loop space as
\bea
g_{Ms,Ns'}(C) &=& g_{MN}(C(s)) \delta(s-t) e(s)
\eea
which does transform covariantly. Then one may write down a reparametrization invariant action for the Yang-Mills term in loop space
\bea
\L &=& \int \frac{ds}{e(s)} \int \frac{ds'}{e(s')} g^{MN}(C(s)) g^{PQ}(C(s')) \tr\(F_{Ms,Ps'}(C) F_{Ns,Qs'}(C)\)
\eea
Here $F(C)$ is the field strength of the gauge potential,
\bea
F(C) &=& \delta A(C) - i A(C) \wedge A(C)
\eea
and $F_{Ms,Ns'}(C)$ are its components, 
\bea
F(C) &=& \int ds \int ds' F_{Ms,Ns'}(C) \delta C^M(s) \wedge \delta C^N(s')
\eea
One may now try to supersymmetrize this Yang-Mills term. To find the supersymmetric action in loop space would be more difficult since we would need to find the correct integration measure on the loop space. 

In Lorentzian signature, we should really think about the nature of the tangent vector of the loops. Our equations in the present paper are not affected by whether this tangent vector is spacelike, timelike or lightlike. Let us compare with an ordinary quantum field theory where the coordinates of the fields can be spacelike, lightlike or timelike with respect to some reference point (the origin). We integrate the field theory Lagrangian over the entire Lorentzian manifold. To understand that the particles see the lightcone structure of spacetime, we may compute a propagator between the field evaluated at two different spacetime points where one of them might be at the origin. Something similar might happen in loop space. The loops in loop space might have a tangent vector that is such that the sign of $\M(s) = \dot{C}^M(s) \dot{C}_M(s)$ can oscillate between being positive, zero and negative as we go around the loop. On the other hand, if we could compute a propagator for a loop field at two different loops, then we may find that it behaves differently depending on how the loops are separated in spacetime. Let us imagine that we could compute the propagator between a pointlike loop and some other loop. Then we may be able to understand what the shape would be of a physical loop by comparing the propagator for various shapes of the loop. When the propagator takes its maximal value, then the second loop may take the shape of a physical loop that has been propagating from the pointlike loop. In this way we may deduce whether $\M(s)$ is positive or zero for the physical loops. The $(2,0)$ theory might be a tensionless string theory. Perhaps our loops could correspond to tensionless strings.

We could also try to find the loop space formulation of the $(2,0)$ theories along the same lines. Then the first problem to overcome for the $(2,0)$ theories would be to find a cohomological formulation. Another problem for the $(2,0)$ theories is a more complicated Fierz identity for the supersymmetry parameter, which is such that the Dirac current $\bar\eps \Gamma^M \eps$ no longer must be lightlike, hence leading to different cases when it is lightlike and when it is not, depending on the choice of $\eps$. So the story becomes more complicated with the $(2,0)$ theories if one tries to uplift them to loop space.

\subsection*{Acknowledgement}
We would like to thank the anonymous referee for asking us critical questions, which helped us to improve the presentation and make the paper more interesting. 
DB was
supported by the 2023 Research Fund of the University of Seoul.
AG was
supported in part by
NRF Grant RS-2023-00208011,  %2020R1A2B5B01001473, 
and by  Basic Science Research Program
through NRF %National Research Foundation 
funded by the Ministry of Education
(2018R1A6A1A06024977).

\appendix

\section*{Appendices}
In these appendices, trace and traceless parts can refer to either one of two situations, $X^I{}_J = X \delta^I_J + \t{X}^I{}_J$ where $\t{X}^I{}_I = 0$ or $X_M = V_M A + U_M B + \t{X}_M$ where $\t{X}_M V^M = 0 = \t{X}_M U^M$. Here tilde quantities will be referred to as traceless parts, and the rest as trace parts.

\section{The Lie derivative $\L_V \eps_I$}\label{sectionA}
We first attempt a direct computation of the Lie derivative,
\bea
\L_V \eps_I &=& V^M \nabla_M \eps_I + \frac{1}{4} \nabla_M V_N \Gamma^{MN} \eps_I\cr
&=& V^M \Gamma_M \eta_I - \frac{1}{2} \Gamma^{MN} \eps_I \bar\eps^J \Gamma_{MN} \eta_J 
\eea
by applying the Fierz identity
\bea
\eps_I \bar\eps^J &=& \frac{1}{8} \delta_I^J V^M \Gamma_M - \frac{1}{24} \Theta_{RST}{}^J{}_I \Gamma^{RST}
\eea
together with the gamma identities
\bea
\Gamma^{UV} \Gamma_M \Gamma_{UV} &=& - 10 \Gamma_M\cr
\Gamma^{UV} \Gamma_{RST} \Gamma_{UV} &=& 6 \Gamma_{RST}
\eea
We then get 
\bea
- \frac{1}{2} \Gamma^{MN} \eps_I \bar\eps^J \Gamma_{MN} \eta_J &=& \frac{5}{8} V^M \Gamma_M \eta_I + \frac{1}{8} \Theta_{RST}{}^J{}_I \Gamma^{RST} \eta_J
\eea
We now like to compare this result with 
\bea
\eps_I \bar\eps^J \eta_J &=& \frac{1}{8} V^M \Gamma_M \eta_I - \frac{1}{24} \Theta_{RST}{}^J{}_I \Gamma^{RST} \eta_J
\eea
By also using that 
\bea
\bar\eps^J \eta_J &=& \frac{1}{12} \nabla_M V^M
\eea
we get relation
\ben
V^M \nabla_M \eps_I - \frac{1}{4} \nabla_M V_N \Gamma^{MN} \eps_I &=& \frac{1}{4} \eps_I \nabla_M V^M\label{eq1}
\een
This relation will be useful for us below. But this direct computation did not go all the way to a final answer. So instead we will now assume that the final answer is of the form 
\ben
\L_V \eps_I &=& \frac{\Omega}{12} \eps_I + \alpha \eps_J R^J{}_I\label{eq2}
\een
and we will now proceed to both prove that this is true as well as fixing the value of the numerical coefficient $\alpha$. By subtracting the two equations (\ref{eq1}) and (\ref{eq2}) from each other, we obtain an algebraic relation for $\eps_I$,
\bea
\frac{1}{2} \nabla_M V_N \Gamma^{MN} \eps_I &=& - \frac{1}{6} \nabla_M V^M \eps_I + \alpha \eps_J R^J{}_I
\eea
By contracting both sides with $\bar\eps^K \Gamma_Q$ we find that the trace part is trivially satisfied, while for the traceless part we get 
\bea
\nabla^M V^N \Theta_{QMN}{}^I{}_J &=& \alpha V_Q R^I{}_J
\eea

The clearest computation we can do in order to fix the coefficient $\alpha$ is 
\bea
U^Q V^N \nabla^M \Theta_{QMN}{}^I{}_J &=& 2 U^Q V^N \bar\eps^I \Gamma_{QMN} \Gamma^M \eta_J \cr
&=& - 8 U^Q V^N \bar\eps^I \Gamma_{QN} \eta_J\cr
&=& 8 U^Q V^N \bar\eps^I \Gamma_{NQ} \eta_J\cr
&=& - 8 \N \bar\eps^I \eta_J\cr
&=& - 8 \N R^I{}_J
\eea
In all steps it being understood that we shall remove any trace part. While this simple computation does not fully prove, it does suggest the following relations
\bea
V^N \nabla^M \Theta_{QMN}{}^I{}_J &=& - 8 V_Q R^I{}_J\cr
U^Q \nabla^M \Theta_{QMN}{}^I{}_J &=& - 8 U_N R^I{}_J
\eea
The first of these relations may also be expressed as
\bea
\nabla^M V^N \Theta_{QMN}{}^I{}_J &=& 8 V_Q R^I{}_J\label{proved}
\eea
and this fixes 
\bea
\alpha &=& 8
\eea
Notice that since we have an identity for the trace part, if we prove that the traceless part corresponds to an identity too, then from (\ref{eq1}) we can prove the equation
\bea
\L_V \eps_I &=& \frac{\Omega}{12} \eps_I + 8 \eps_J R^J{}_I
\eea
We will now proceed to prove equation (\ref{proved}). We compute
\bea
- 4 R^I{}_J \Psi^{AJ}_Q &=& - 4 \(\bar\eps^I \eta_J - \frac{1}{2} \delta^I_J \bar\eps^K \eta_K\) \bar\eps^J \Gamma_Q \Psi^A
\eea
by first computing the first term,
\bea
I &=& - 4 \bar\eps^I \eta_J \bar\eps^J \Gamma_Q \Psi^A
\eea
using 
\bea
\eta_J \bar\eps^I &=& \frac{1}{48} \nabla_N V^N + \frac{1}{16} \Gamma^{MN} \nabla_M V_N
\eea
We then get
\bea
I &=& - \frac{1}{12} \nabla_N V^N \Psi_Q^{AI} - \frac{1}{4} \Psi^{AI,M} \(\nabla_M V_N - \nabla_N V_M\)\cr
&& - \frac{1}{4} \Psi^{AI,MN}{}_Q \nabla_M V_N\cr
&=& - \frac{1}{6} V_Q \nabla_N V^N \Psi^{AI} - \frac{1}{4} \Psi^{AI,MN}{}_Q \nabla_M V_N
\eea
Next we compute the second term
\bea
II &=& \frac{1}{2} \delta^I_J \bar\eps^K \eta_K V_Q \Psi^{AJ} \cr
&=& \frac{1}{6} V_Q \nabla_N V^N \Psi^{AI}
\eea
By considering the sum, we obtain the relation 
\bea
- 4 R^I{}_J \Psi^{AJ}_Q &=& - \frac{1}{4} \Psi^{AI,MN}{}_Q \nabla_M V_N
\eea
Now let us contract with $U^Q$ on both sides and use
\bea
\Psi_{MN}^{AI} &=& 2 \Theta_{MN}{}^I{}_J \Psi^{AJ}
\eea
Then we get
\bea
- 4 R^I{}_J \Psi^{AJ} \N &=& - \frac{1}{2} \Theta^{MN,I}{}_J \Psi^{AJ} \nabla_M V_N
\eea
and so
\bea
\Theta^{MN,I}{}_J \nabla_M V_N &=& 8 \N R^I{}_J 
\eea

\section{The operators $\delta$ and $\delta^{\dag}$ in loop space}\label{dagger}
The loop space differential $\delta$ and its adjoint $\delta^{\dag}$ are in one-to-one correspondence with corresponding spacetime differential $d$ and its adjoint $d^{\dag}$. Let us consider the zero-form in loop space
\bea
\Lambda &=& \int ds \lambda_M \dot{C}^M
\eea
Acting with $\delta$, we get a one-form in loop space 
\bea
\delta \Lambda &=& \int ds \(\partial_M \lambda_N - \partial_N \lambda_M\) \delta C^M \dot{C}^N
\eea
For the one-form in loop space
\bea
A &=& \int ds B_{MN} \delta C^M \dot{C}^N
\eea
we get the two-form in loop space by acting on it with $\delta$
\bea
\delta A &=& \int ds \frac{1}{2} \(\partial_P B_{MN} + \partial_M B_{NP} + \partial_N B_{PM}\) \delta C^M \wedge \delta C^N \dot{C}^P
\eea
The small computations to obtain these results can be found in \cite{Freund:1981qw}. 

We would now like to define the adjoint operator $\delta^{\dag}$ acting on the one-form $A$ to produce a zero-form. We have another zero-form $\Lambda$, which we shall, in this context, think of as a test function in loop space. Then we could imagine taking their inner product in loop space and define $\delta^{\dag}$ as follows,
\bea
(\delta^{\dag} A,\Lambda) &=& (A,\delta \Lambda)
\eea
The only problem is that the inner product in loop space is hard to define. But we may observe the correspondence between spacetime and loop space fields, which makes it natural to define the inner product in loop space as the inner product of the corresponding spacetime fields,
\bea
(A, \delta \Lambda) &=& (B,d\lambda)
\eea
Then we may perform the standard manipulations on the spacetime side and finally translate the result there, back to loop space. Thus the computation would be
\bea
(A,\delta \Lambda) = (B,d\lambda) = (d^{\dag} B,\lambda) = (\delta^{\dag} A,\Lambda)
\eea
which leads to 
\bea
\delta^{\dag} A &=& - \int ds \nabla^M B_{MN} \dot{C}^N
\eea

\section{The hypermultiplet field $\Up_Q^{AI}$}\label{UpQAI}
In this section our aim is to study the supersymmetry variation of the cohomological field 
\bea
\Up^{AI}_Q &=& \bar\eps^I \Gamma_Q \Gamma_M \psi^A U^M
\eea
in the hypermultiplet and how this is uplifted to loop space. First let us notice that this field has only four transverse components since $V^Q \Up_Q^{AI} = 0$ and $U^Q \Up_Q^{AI} = 0$, which is almost manifest. We just need to notice that $V^Q \bar\eps_I \Gamma_Q = 0$ and $\Gamma_Q\Gamma_M U^Q U^M = 0$. When we uplift its supersymmetry variation to loop space, then all the complications come from the geometry. The nonabelian generalization is trivial. So let us assume that the gauge group is abelian. Our starting point is the innocent-looking supersymmetry variation
\bea
\delta \psi^A &=& - \Gamma^M \eps_I \partial_M q^{AI} - 4 \eta_I q^{AI}
\eea
While this may look innocent, it turns out to be far from trivial to uplift this supersymmetry variation to the loop space. A short computation shows that the induced supersymmetry variation of $\Up^{AI}_Q$ becomes
\bea
\delta \Up^{AI}_Q &=& - \frac{1}{2} V_Q \L_U q^{AI} - \frac{1}{2} U_Q \L_V q^{AI} + \frac{\N}{2} \nabla_Q q^{AI}\cr
&& + \Theta_{QMN}{}^I{}_J U^N \nabla^M q^{AJ}\cr
&& - 4 U^N \bar\eps^I \Gamma_Q \Gamma_N \eta_J q^{AJ}\label{Hett}
\eea
By looking at this variation, it is easy to see that 
\bea
V^Q \delta \Up_Q^{AI} &=& 0\cr
U^Q \delta \Up_Q^{AI} &=& 0
\eea
Next, we claim that the supersymmetry variation above can be recast in the following form
\bea
\delta \Up_Q^{AI} &=& - \frac{1}{2} V_Q \(\L_U q^{AI} + \frac{\Omega^{\vee}}{3} q^{AI}\)\cr
&& - \frac{1}{2} U_Q \(\L_V q^{AI} + \frac{\Omega}{3} q^{AI}\)\cr
&& + \frac{1}{2} \nabla_Q \(\N q^{AI}\)\cr
&& + \Theta_{QMN}{}^I{}_J \nabla^M \(U^N q^{AJ}\)\label{Htva}
\eea
If we further define 
\bea
Q^{AI} &=& \frac{\N}{2} q^{AI}\cr
Q^{AI}_{MN} &=& \Theta_{QMN}{}^I{}_J U^Q q^{AJ}
\eea
then 
\bea
\delta \Up_Q^{AI} &=& \nabla_Q Q^{AI} - \frac{1}{\N} V_Q \L_U Q^{AI} - \frac{1}{\N} U_Q \L_V Q^{AI} \cr
&& + \nabla^M Q_{QM}^{AI} - \(\nabla^M \Theta_{QMN}{}^I{}_J\) U^N q^{AJ}
\eea
The purpose of the last term becomes clear if we compute the $U$ and $V$ traces. Namely, the first line is manifestly $U$ and $V$ traceless, and hence the purpose of the last term is to remove the trace parts from $\nabla^M Q_{QM}^{AI}$, which we may indicate by a tilde,
\bea
\t{\nabla^M Q_{QM}^{AI}} &=& \nabla^M Q_{QM}^{AI} - \(\nabla^M \Theta_{QMN}{}^I{}_J\) U^N q^{AJ}
\eea
Then the resulting supersymmetry variation reads 
\bea
\delta \Up_Q^{AI} &=& \nabla_Q Q^{AI} - \frac{1}{\N} V_Q \L_U Q^{AI} - \frac{1}{\N} U_Q \L_V Q^{AI} + \t{\nabla^M Q_{QM}^{AI}} 
\eea
This is now of a form that we can uplift to loop space. We define the loop space fields
\bea
\Up^{AI} &=& \int ds \dot{C}^Q \Up^{AI}_Q\cr
\Phi^{AI} &=& \int ds \frac{2\dot{C}^Q V_Q}{\N} Q^{AI}\cr
\Phi^{\vee,AI} &=& \int ds \frac{2\dot{C}^Q U_Q}{\N} Q^{AI}\cr
Q_1^{AI} &=& \int ds Q^{AI}_{MN} \dot{C}^M \delta C^N
\eea
Then our final result for the supersymmetry variation in loop space reads
\bea
\delta \Up^{AI} &=& - \frac{1}{2} \L_U \Phi^{AI} - \frac{1}{2} \L_V \Phi^{\vee,AI} + \delta^{\dag} Q_1^{AI}
\eea
where we define a loop space version of the adjoint of the differential operator $\delta$ as
\bea
\delta^{\dag} Q_1^{AI} &=& \int ds \dot{C}^Q \t{\nabla^M Q_{QM}^{AI}}
\eea
Since $\Up_Q^{AI}$ lives in the transverse four-dimensional space due to the constraints $V^Q \Up_Q^{AI} = 0$ and $U^Q \Up_Q^{AI}$, the adjoint $\delta^{\dag}$ shall be defined with respect to this transverse space through our correspondence between spacetime and loop space as explained in the previous section. It is also important to notice that we do not restrict the tangent vector $\dot{C}^M$ of the loop to the transverse four-dimensional space. Our loops always live in six-dimensional spacetime and $\dot{C}^M$ is arbitrary, and it can be spacelike, timelike or lightlike and even oscillate between these various types as we go around the loop. 

We may also notice that 
\bea
\t{\nabla_Q Q^{AI}} &=& \nabla_Q Q^{AI} - \frac{1}{\N} V_Q \L_U Q^{AI} - \frac{1}{\N} U_Q \L_V Q^{AI}
\eea
so we may write
\bea
\delta \Up_Q^{AI} &=& \t{\nabla_Q Q^{AI}} + \t{\nabla^M Q_{QM}^{AI}} 
\eea
Curiously the uplift to loop space of the first two terms can now be expressed as 
\bea
\int ds \dot{C}^M \t{\nabla_M Q^{AI}} &=& - \frac{1}{2} \L_U \Phi^{AI} - \frac{1}{2} \L_V \Phi^{\vee,AI}
\eea
It is almost a total derivative, but a total derivative would vanish when integrated over the loop. The only reason why this does not vanish is because of the projection to the transverse four-dimensional space. 

\subsection{Equivalence of the two expressions for $\delta \Up_Q^{AI}$}
For the two expressions for the supersymmetry variation in (\ref{Hett}) and (\ref{Htva}) to agree, we need the following two identities,
\bea
\Theta_{QMN}{}^I{}_J \nabla^M U^N &=& - 4 \t{\bar\eps^I \Gamma_N \Gamma_M \eta_J} U^M\cr
\frac{\Omega^{\vee}}{6} V_Q  + \frac{\Omega}{6} U_Q  - \frac{1}{2} \nabla_Q \N &=& 2 \bar\eps^I \Gamma_Q \Gamma_M \eta_I U^M
\eea
The second of these identities can be easily shown as follows. The right-hand is 
\bea
2 \bar\eps^I \Gamma_Q \nabla_M \eps_I U^M &=& \nabla_M V_Q U^M
\eea
Next we expand 
\bea
-\frac{1}{2} \nabla_Q \N &=&  - \frac{1}{2} U^M \nabla_Q V_M - \frac{1}{2} V^M \nabla_Q U_M
\eea
Applying the conformal Killing vector equations on both terms, that is, for both $V_M$ and $U_M$, we get
\bea
- \frac{1}{2} \nabla_Q \N &=&  - \frac{\Omega}{6} U_Q - \frac{\Omega^{\vee}}{6} V_Q + \frac{1}{2} U^M \nabla_M V_Q + \frac{1}{2} V^M \nabla_M U_Q
\eea
and by finally applying $[V,U]_Q = 0$ we see that the left-hand side equals the right-hand side. 

The first identity can be recast in the form
\bea
\frac{1}{4} \t{\bar\eps^I \Gamma_Q \Gamma_{MN} \eps_J} \nabla^M U^N + \t{\bar\eps^I \Gamma_Q \Gamma_M \eta_J} U^M &=& 0
\eea
or, equivalently,
\bea
\t{\bar\eps^I \Gamma_Q \L_U \eps_J} &=& 0\label{Qu}
\eea
This equation will be satisfied if  
\bea
\L_U \eps_I &=& \frac{\Omega^{\vee}}{12} \eps_I 
\eea
While we have not been able to prove this equation, we can see that it is consistent with 
\bea
\L_U V_M &=& \frac{\Omega^{\vee}}{3} V_M
\eea
which follows from assuming that $U$ and $V$ are two commuting conformal Killing vectors.

\subsection{Mini loop space}
In mini loop space, we restrict the loops such that $\dot{C}^M = v^M$ where $v^M$ is assumed to be a spatial Killing vector. We can then make some calculations more precise, where we may use additional inputs coming from constraining fields such that $\L_v$ vanishes on all spacetime fields and on all geometric quantities, such as $\L_v V_M = 0$ and so on. Our loop space fields become in the mini loop space
\bea
\Up^{AI} &=& v^M \Up_M^{AI}\cr
\Phi^{AI} &=& v^M V_M q^{AI}\cr
\Phi^{\vee,AI} &=& v^M U_M q^{AI}
\eea
by taking the range of integration such that $\int ds = 1$. In the mini loop space we can compute the Lie derivative of $\Phi^{\vee,AI}$ as
\bea
\L_V \Phi^{\vee,AI} &=& V^M \nabla_M \Phi^{\vee,AI}\cr
&=& V^M \nabla_M \(v^N U_N q^{AI}\)\cr
&=& (\L_V v_N) U^N q^{AI} + v_N U^N \L_V q^{AI}
\eea
Now we use that $\L_v V_M = 0$ together with $\nabla_M v_N + \nabla_N v_M = 0$, which  implies that $V^P \nabla_P v_M = v^P \nabla_P V_M$ and therefore
\bea
\L_V v_N &=& v^P \(\nabla_P V_N + \nabla_N V_P\)
\eea
Finally by using that $V_M$ is a conformal Killing vector, we get
\bea
\L_V \Phi^{\vee,AI} &=& v^N U_N \(\L_V q^{AI} + \frac{\Omega}{3} q^{AI}\)
\eea
To get the corresponding result for $\L_U \Phi^{AI}$, we need instead to use the conformal Killing equation for $U^M$. Only then do we have 
\bea
\L_U \Phi^{AI} &=& v^N V_N \(\L_U q^{AI} + \frac{\Omega^{\vee}}{3} q^{AI}\)
\eea
Let us now compute 
\bea
- \nabla^M \Theta_{QMN}{}^I{}_J v^Q U^N &=& - 2 \t{\bar\eps^I \Gamma_{QMN} \Gamma^M \eta_J} v^Q U^N\cr
&=& 8 \t{\bar\eps^I \Gamma_{QN} \eta_J} v^Q U^N\cr
&=& 8 R^I{}_J v^N U_N - 8 \bar\eps^I \Gamma_N \Gamma_Q \eta_J v^Q U^N\cr
&=& 8 R^I{}_J v^N U_N + 2 \Theta_{NPQ}{}^I{}_J \nabla^P v^Q U^N
\eea 
Here the second term on the last line vanishes, as the following computation shows,
\bea
\Theta_{PQN,IJ} \nabla^P v^Q &=& \t{\bar\eps_I \Gamma_{PQ} \Gamma_N \eps_J} \nabla^P v^Q + \t{\bar\eps_I \Gamma_N \Gamma_{PQ} \eps_J} \nabla^P v^Q\cr
&\sim& \t{\bar\eps_I \Gamma_N \Gamma_M \eta_J} v^M - \t{\bar\eta_I \Gamma_M \Gamma_N \eps_J} v^M\cr
&=& v^M \t{\bar\eps_I \Gamma_N \nabla_M \eps_J} + v^M \t{\nabla_M \bar\eps_I \Gamma_N \eps_J}\cr
&=& v^M \nabla_M \(\t{\bar\eps_I \Gamma_N \eps_J}\)
\eea
This vanishes since $\bar\eps_I \Gamma_N \eps_J$ is antisymmetric in $I$ and $J$, which means that it has no traceless part. As a result, we have the relation
\bea
v^Q U^N \nabla^M \Theta_{QMN}{}^I{}_J &=& - 8 v^Q U_Q R^I{}_J
\eea
that we obtained by just using $\L_v \eps_I = 0$.

\section{The tensor $R^I{}_J$ is covariantly constant}\label{RIJ}
We define 
\bea
R^I{}_J &=& \bar\eps^I \eta_J - \frac{1}{2} \delta^I_J \bar\eps^K \eta_K
\eea 
which is defined such that it is traceless,
\bea
R^I{}_I &=& 0
\eea
Alternatively we may define 
\bea
R_{IJ} &=& \bar\eps_I \eta_J + \frac{1}{2} \ep_{IJ} \bar\eps^K \eta_K
\eea
such that 
\bea
\ep^{IJ} R_{IJ} &=& 0
\eea
Here $\ep^{IJ} \ep_{IJ} =- 2$ and $\bar\eps^I = \bar\eps_I \ep^{IJ}$. To show that $R_{IJ}$ is covariantly constant, it is enough to show that $\nabla_M \(\bar\eps_I \eta_J\)$ is antisymmetric in $I$ and $J$ since the antisymmetric part is subtracted when we compute $\nabla_M R_{IJ}$. We have
\ben
\nabla_M \(\bar\eps_I \eta_J\) &=& - \bar\eta_I \Gamma_M \eta_J + K_{MN} \bar\eps_I \Gamma^N \eps_J\label{covcon}
\een
where
\bea
K_{MN} &=& \frac{R}{80} g_{MN} - \frac{1}{8} R_{MN} 
\eea
Here $g_{MN}$ is the metric tensor and $R_{MN}$ is the Ricci curvature of the six-manifold. Now the right-hand side of (\ref{covcon}) is antisymmetric in $I$ and $J$ as a result of the Majorana conditions of both $\eps_I$ and $\eta_I$.

\section{Lie derivatives in loop space}\label{LieDer}
For a vector field $v^M(x)$ in spacetime we associate a corresponding vector field in loop space
\bea
v^{Ms}(C) &=& v^M(C(s)) ds
\eea
The Lie derivative of a function in loop space
\bea
f(C) &=& \int ds f_M(C(s)) \dot{C}^M(s)
\eea
is given by
\bea
\L_v f(C) &=& v^{Ms} \partial_{Ms} f(C)\cr
&=& \int ds v^M(C(s)) \frac{\delta f(C)}{\delta C^M(s)}
\eea
The functional derivative is defined as
\bea
\delta f(C) &=& \int ds \frac{\delta f(C)}{\delta C^M(s)} \delta C^M(s)
\eea
so in this case it is given by
\bea
\frac{\delta f(C)}{\delta C^N(s)} &=& \(\partial_M f_N - \partial_N f_M\) \dot{C}^M
\eea
and so
\bea
\L_v f(C) &=& \int ds \(\partial_N f_M - \partial_M f_N\) \dot{C}^M v^N\cr
&=& \int ds \(v^N \partial_N f_M \dot{C}^M - \frac{d f_N}{ds} v^N\)\cr
&=& \int ds \(v^N \partial_N f_M \dot{C}^M + \frac{d v^N}{ds} f_N\)\cr
&=& \int ds \(v^N \partial_N f_M + \partial_M v^N f_N\) \dot{C}^M\cr
&=& \int ds \(\L_v f_M\) \dot{C}^M
\eea

Let us next compute the Lie derivative of a one-form in loop space
\bea
f(C) &=& \int ds f_{MN} \dot{C}^M \delta C^N
\eea
The components of this one-form are
\bea
f_{Ns} &=& f_{MN}(C(s)) \frac{dC^M}{ds}
\eea
This suggests that we shall compute the Lie derivative as
\bea
\L_v f_{Ms} &=& v^{Nt} \partial_{Nt} f_{Ms} + \(\partial_{Ms} v^{Nt}\) f_{Nt}\cr
&=& \int dt v^N \frac{\delta f_{Ms}}{\delta C^N(t)} + \int dt \frac{\delta v^N(C(t))}{\delta C^M(s)} f_{PN} \frac{dC^P(t)}{dt} 
\eea
We have
\bea
\frac{\delta f_{Ms}}{\delta C^N(t)} &=& \partial_N f_{PM} \dot{C}^P \delta(s-t) - \frac{d}{dt} \[f_{NM}(C(t)) \delta(t-s)\] \cr
\frac{\delta v^N(C(t))}{\delta C^M(s)} &=& \partial_M v^N \delta(s-t)
\eea
Then we get
\bea
\L_v f_{Ms} &=& \L_v f_{PM} \frac{dC^P}{ds} 
\eea
where
\bea
\L_v f_{PM} &=& v^N \partial_N f_{PM} + \partial_P v^N f_{NM} + \partial_M v^N f_{PN}
\eea

We define 
\bea
A(C) &=& \oint ds B_{MN}(C(s)) \dot{C}^M(s) \delta C^N(s)
\eea
For the scalar field we define a corresponding loop space field as
\bea
\Phi(C) &=& \oint ds V_M(C(s)) \dot{C}^M(s) \phi(C(s)) 
\eea
Under an infinitesimal variation of the loop, this field has the variation
\bea
\delta \Phi(C) &=& \oint ds \(\delta C^Q \dot{C}^M \phi \(\partial_Q V_M - \partial_M V_Q\) - \delta C^Q V_Q \partial_M \phi + \delta C^Q V_M \dot{C}^M \partial_Q \phi\)
\eea
The functional derivative is therefore
\bea
\frac{\delta \Phi(C)}{\delta C^Q(s)} &=& \(\partial_Q V_M - \partial_M V_Q\) \dot{C}^M \phi - V_Q \dot{C}^M \partial_M \phi + V_M \dot{C}^Q \partial_Q \phi
\eea
We define the Lie derivative as
\bea
\L_V \Phi(C) &=& \oint ds V^Q(C(s)) \frac{\delta \Phi(C)}{\delta C^Q(s)}
\eea
By using 
\bea
V^Q V_Q &=& 0\cr
\nabla_M V_N + \nabla_N V_M &=& \frac{1}{3} g_{MN} \nabla_P V^P
\eea
this Lie derivative becomes
\bea
\L_V \Phi(C) &=& \oint ds V_M \dot{C}^M \(V^Q \partial_Q \phi + \frac{1}{3} \nabla_M V^M \phi\)
\eea

Let us finally compute
\bea
\L_U \Phi(C) &=& \int ds \dot{C}^M \(V_M U^P \nabla_P \phi + \(U^N \nabla_N V_M + V^N \nabla_M U_N\) \phi\)
\eea 
To proceed here, we need to assume that $U$ is a conformal Killing vector. Only then can we rewrite the prefactor in the last term as
\bea
V^N \nabla_M U_N &=& \frac{1}{3} V_M \nabla_P U^P - V^N \nabla_N U_M
\eea
and then we find the combination
\bea
U^N \nabla_N V_M - V^N \nabla_N U_M &=& 0
\eea
leading to the result
\bea
\L_U \Phi(C) &=& \int ds \dot{C}^M V_M \(U^P \nabla_P \phi + \frac{1}{3} \nabla_P U^P \phi\)
\eea

\section{Properties of $\Theta_{MNP}{}^I{}_J$} \label{APPF}
We have that 
\bea
\Theta_{MNP}{}^I{}_J V^P &=& 0
\eea
Let us now define 
\bea
%\label{F2}
\Theta_{MN}{}^I{}_J &=& \Theta_{MNP}{}^I{}_J U^P
\eea
We will now show that all information about $\Theta_{MNP}{}^I{}_J$ can be recovered from $\Theta_{MN}{}^I{}_J$.

We begin by noting that 
\bea
\eps_J &=& - \frac{1}{6\N} \Gamma^{MN} \eps_I \Theta_{MN}{}^I{}_J 
\eea
that we can derive using 
\bea
\eps_I \bar\eps^I &=& \frac{1}{4} V^Q \Gamma_Q\cr
\Gamma^{MN} \Gamma^Q \Gamma_{MNP} &=& - 20 \delta^Q_P + 4 \Gamma_P{}^Q\cr
U_M V_N \Gamma^{MN} \eps_J &=& - \N \eps_J
\eea
We next use this formula to expand 
\bea
\Theta_{RST}{}^K{}_J &=& \bar\eps^K \Gamma_{RST} \eps_J\cr
&=& - \frac{1}{6\N} \bar\eps^K \Gamma_{RST} \Gamma^{MN} \eps_I \Theta_{MN}{}^I{}_J
\eea
We use
\bea
\Gamma_{RST}\Gamma^{MN} &=& - 6 \delta_{RS}^{MN} \Gamma_T + \Gamma_{RST}{}^{MN} + 6 \delta^{[M}_{[R} \Gamma_{ST]}{}^{N]}
\eea
Then the right-hand side becomes
\bea
\frac{1}{2\N} V_T \Theta_{RS}{}^K{}_J - \frac{1}{12 \N} \eps_{RST}{}^{MNP} \Theta_{MN}{}^K{}_J V_P - \frac{1}{\N} \Theta_{ST}{}^{NK}{}_I \Theta_{RN}{}^I{}_J
\eea
We analyze the third term, using
\bea
\Gamma_{ST}{}^N \Gamma^Q \Gamma_{RNP} &=& -2 \delta^Q_R \Gamma_{STP} + 2 g_{PR} \Gamma^Q{}_{ST} - 2 \delta^Q_P \Gamma_{RST}
\eea
and 
\bea
\eps_I \bar\eps^I &=& \frac{1}{4} \Gamma^Q V_Q
\eea
Then 
\bea
 - \frac{1}{\N} \Theta_{ST}{}^{NK}{}_I \Theta_{RN}{}^I{}_J &=& \frac{1}{2\N} \Theta_{ST}{}^K{}_J V_R + \frac{1}{2} \Theta_{RST}{}^K{}_J
 \eea
So we get 
\bea
\Theta_{RST}{}^K{}_J &=& \frac{1}{2\N} V_T \Theta_{RS}{}^K{}_J - \frac{1}{12 \N} \eps_{RST}{}^{MNP} \Theta_{MN}{}^K{}_J V_P \cr
&& + \frac{1}{2\N} \Theta_{ST}{}^K{}_J V_R + \frac{1}{2} \Theta_{RST}{}^K{}_J
\eea
This is an equation that gives us the answer
\bea
\Theta_{RST}{}^K{}_J &=& \frac{2}{\N} \Theta_{RS}{}^K{}_J V_T - \frac{1}{6 \N} \eps_{RST}{}^{MNP} \Theta_{MN}{}^K{}_J V_P
\eea
We can also write this answer as
\bea
\Theta_{RST}{}^K{}_J &=& \frac{3}{2\N} \Theta_{RS}{}^K{}_J V_T - \frac{1}{4\N} \eps_{RST}{}^{MNP} \Theta_{MN}{}^K{}_J V_P\cr
&+& \frac{1}{2\N} \Theta_{RS}{}^K{}_J V_T + \frac{1}{12 \N} \eps_{RST}{}^{MNP} \theta_{MN}{}^K{}_J V_P
\eea
We also know that 
\bea
\Theta_{RST} &=& \Theta_{RST}^-
\eea
From these results we learn two things. First that $\Theta_{MNP}{}^I{}_J$ can be constructed out of $\Theta_{MN}{}^I{}_J$ as
\bea
\Theta_{RST}{}^K{}_J &=& \frac{3}{2\N} \Theta_{RS}{}^K{}_J V_T - \frac{1}{4\N} \eps_{RST}{}^{MNP} \Theta_{MN}{}^K{}_J V_P
\eea
and, second, that $\Theta_{MN}{}^I{}_J$ satisfies its own selfduality relation
\bea
0 &=& \Theta_{RS}{}^K{}_J + \frac{1}{2 \N} \eps_{RS}{}^{TMNP} \Theta_{MN}{}^K{}_J V_P U_T
\eea

\section{Conformally Einstein manifolds}\label{ConfEin}
If on a Lorentzian six-manifold we have two commuting symplectic Majorana nonchiral spinors $\E_I$ for $I = 1,2$ such that $\bar\E^I \E_I$ is nowhere vanishing and satisfy the conformal Killing spinor equation 
\bea
\nabla_M \E_I &=& \Gamma_M \Pi_I
\eea
where $\Pi_I$ is some other symplectic Majorana nonchiral spinor, then the six-manifold has to be conformally equivalent to an Einstein manifold. 

To show this for nonchiral spinors, we follow a similar proof in \cite{baumsbook} for Riemannian manifolds. By the symplectic Majorana condition we have that $- i\bar\E^I \E_I$ is real. We may then without restriction assume that $-i \bar\E^I \E_I > 0$ everywhere. Under a Weyl transformation of the metric the spinors transform as $\E_I \rightarrow e^{\sigma/2} \E_I$. We can make a Weyl transformation such that  
\bea
\bar\E^I \E_I &=& i
\eea
By acting with one derivative on this relation, we get
\bea
\bar\E^I \Gamma_M \Pi_I &=& 0
\eea
By acting again with another derivative we get
\bea
- \bar\Pi^I \Gamma_N \Gamma_M \Pi_I + K_{MN} \bar\E^I \E_I &=& 0
\eea
where 
\bea
K_{MN} &=& \frac{1}{8} \(\frac{R}{10} g_{MN} - R_{MN}\)
\eea
By noting the normalization of $\E_I$ and the fact that $\bar\Pi^I \Gamma_{MN} \Pi_I = 0$ we get
\bea
K_{MN} &=& - i g_{MN} \bar\Pi^I \Pi_I
\eea
This is equivalent to 
\bea
R_{MN} &=& \(\frac{R}{10} + 8 i \bar\Pi^I \Pi_I\) g_{MN}
\eea
Taking the trace of both sides we get 
\bea
R &=& \frac{3R}{5} + 48 i \bar\Pi^I \Pi_I
\eea
from which we conclude that $\Pi_I$ satisfies the normalization condition
\bea
\bar\Pi^I \Pi_I &=& - \frac{i R}{120}
\eea
Then we get  
\bea
R_{MN} &=& \(\frac{R}{10} + \frac{R}{15}\) g_{MN}\cr
&=& \frac{R}{6} g_{MN}
\eea
which shows that the six-manifold is an Einstein manifold for this particular choice of metric. As a corollary, the six-manifold is conformally equivalent to an Einstein manifold if $\E_I$ are two conformal Killing spinors that are such that $-i \bar{\E}^I \E_I$ is nowhere vanishing but not necessarily unit normalized. 

If $\eps_I$ is chiral, so that $\Gamma \eps_I = - \eps_I$, then $\bar\eps^I \Gamma = + \bar\eps^I$ and we get $\bar\eps^I \eps_I = 0$. For two chiral conformal Killng spinors, by following through the above reasoning, we see that the manifold does not have to be related to an Einstein manifold in any way. On the other hand, for $\E_I$ nonchiral, the corresponding Dirac current $\bar\E^I \Gamma_M \E_I$ does not have to be lightlike. We shall therefore pick the chiral components from $\E_I$ and $\Pi_I$ that generate the $(1,0)$ supersymmetries and from these chiral components we construct the Dirac currents $U$ and $V$ that will both be lightlike. This does not change the fact that nonchiral spinors $\E_I$ and $\Pi_I$ would exist on the manifold. These Weyl components are also conformal Killing spinors subject to $\Gamma \eps_I = - \eps_I$ and $\Gamma \eta_I = + \eta_I$. For these Weyl spinors and their associated Dirac currents, we may compute their Lie bracket $[U,V]$ by assuming that we have made a proper Weyl transformation such that the manifold is an Einstein manifold where $U$ is the Dirac current of $\eta_I$. Then a short computation shows that their Lie bracket
\bea
[U,V]^M &=& U^N \nabla_N V^M - V^N \nabla_N U^M
\eea
is given by 
\bea
[U,V]^M &=& 2 \bar\eps^J \Gamma^M \Gamma_N \eta_J \(U^N - K^{NP} V_P\) + 4 \bar\eps^J \eta_J K^{MN} V_N
\eea
Here we have used the symplectic Majorana condition. We next use the identity
\ben
\nabla_M \Omega &=& 12 \(K_{MN} V^N - U_M\)\label{Om}
\een
that follows from eq (\ref{covcon}), where as before $\Omega = \nabla_M V^M$. Then we get
\bea
[U,V]^M &=& \frac{1}{12} \nabla^N V^M \partial_N \Omega + \frac{1}{3} \Omega K^{MN} V_N
\eea
But since we have an Einstein manifold we have $K_{MN} = - \frac{R}{120} g_{MN}$ and
\bea
[U,V]^M &=& \frac{1}{12} \(\nabla^N V^M \partial_N \Omega - \frac{R}{30} \Omega V^M \)
\eea
We can always make another Weyl transformation that brings $\Omega = 0$. Then $[U,V] = 0$ for that metric. But the Lie bracket, which can be expressed as 
\bea
[U,V]^M = U^N \partial_N V^M - V^N \partial_N U^M
\eea
does not depend on the metric. (We assume that there is no torsion.) Nor does it get affected by a Weyl transformation since both $U^M$ and $V^M$ are Weyl invariant vector fields. That means that since $[U,V] = 0$ for one particular Weyl transformed metric, the Lie bracket vanishes for all metrics that are related to that particular metric by a Weyl transformation. As a corollary we see that 
\ben
\nabla^N V^M \partial_N \Omega &=& \frac{R \Omega}{30} V^M \label{curiousid}
\een
must be an identity that holds for any conformally Einstein manifold. We may also check explicitly its consistency by contracting with $U_M$ and $V_M$ respectively. If we contract both sides with $U_M$ and apply eq (\ref{Om}) in the form $\partial_M \Omega = \frac{R}{10} V_M - U_M$, then we get $U_N V_M \nabla^N V^M \frac{R}{10} = \frac{\N R \Omega}{30}$, which is a true identity, as one can show by using the fact that $V$ is a lightlike conformal Killing vector. Again, if we contract (\ref{curiousid}) with $V_M$ we get $0 = 0$ by using that $V$ is lightlike, so again it is consistent. 

In conclusion, we have shown that a lightlike conformal Killing vector $U$ exists and that it always commutes with $V$ if we assume that there exists at least two symplectic Majorana nonchiral conformal Killing spinor $\E_I$ such that $\bar\E^I\E_I$ is nowhere vanishing. 

As a corollary, we have the identity
\ben
\L_V \Omega^{\vee} &=& \L_U \Omega\label{ovee}
\een
The proof of this identity uses the conformal Killing vector equations for both $U$ and $V$ as well as the fact that $[U,V] = 0$, 
\bea
\L_U \Omega &=& \L_U \(\frac{1}{2} g^{MN} \L_V g_{MN}\)\cr
&=& \frac{1}{2} \L_U g^{MN} \L_V g_{MN} + \frac{1}{2} g^{MN} \L_U \L_V g_{MN}\cr
&=& \frac{1}{2} \L_U g^{MN} \L_V g_{MN} + \frac{1}{2} g^{MN} \L_V \L_U g_{MN}\cr
&=& \frac{1}{2} \(- \frac{\Omega^{\vee}}{3} g^{MN}\) \frac{\Omega}{3} g_{MN} + \frac{1}{2} \L_V \(\frac{\Omega^{\vee}}{3} g_{MN}\) g^{MN}\cr
&=& \L_V \Omega^{\vee}
\eea
From this identity we can see that we can always find a Weyl transformation such that we get both $\Omega = 0$ and $\Omega^{\vee} = 0$, by solving the equations $6\L_V \sigma + \Omega = 0$ and $6\L_U \sigma + \Omega^{\vee} = 0$. That it is possible to solve these equations is because they imply $6\L_U \L_V \sigma + \L_U \Omega = 0$ and $6\L_V \L_U \sigma + \L_V \Omega^{\vee} = 0$ and these equations are mutually consistent since they are equivalent with each other by noting that $[\L_U,\L_V] = 0$. 

Finally by assuming that both $U$ and $V$ have been brought to be Killing vectors, then we may form a timelike Killing vector as $T^M = U^M + V^M$ if we assume that $U^M V_M < 0$. We may then write the metric as $g_{MN} = -  \frac{T_M T_N}{T^2} + \t{g}_{MN}$ where $T^M \t{g}_{MN} = 0$ and $T^2 = - g_{MN} T^M T^M$. We define the zeroth component of the vielbein of this metric as $e^0 = \frac{T_M}{T} dx^M$. The hermitian conjugate of the 6d gamma matrices when they are expressed in tangent space indices, is $(\Gamma^A)^{\dag} = \Gamma^T \Gamma^A \Gamma^T$ where we define $\Gamma^T = \Gamma^M \frac{T_M}{T}$. By converting into curved space indices, which we do by contracting with the inverse vielbein $e_A^M$, we find that the hermitian conjugate rule for those gamma matrices is given by $(\Gamma^M)^{\dag} = \Gamma^T \Gamma^M \Gamma^T$. We can now show that the projection operator is hermitian as follows. We only need to show that $U_M V_N \Gamma^{MN}$ is hermitian. Its hermitian conjugate is $U_M V_N (\Gamma^{MN})^{\dag} = U_M V_N \Gamma^T \Gamma^{MN} \Gamma^T$. Let us compute the difference, multiplied with $\Gamma^T$ from the right,
\bea
\(U_M V_N \Gamma^{MN} - U_M V_N (\Gamma^{MN})^{\dag}\) \Gamma^T &=& U_M V_N \(\Gamma^T \Gamma^{MN} + \Gamma^{MN} \Gamma^T\)\cr
&=& 2 U_M V_N T_P \Gamma^{MNP}\cr
&=& 2 U_M V_N (U_P + V_P) \Gamma^{MNP}\cr
&=& 0
\eea
showing that the projection operators $P_{\pm}$ constructed out of $U$ and $V$ are hermitian. From this we conclude that the vector field $U$ that we here have constructed as the Dirac current of $\eta_I$ must be the same vector field $U$ that we constructed as a vector field that enabled us to form the Weyl projection operators $P_{\pm}$ in the main text. 

The norm $- i \bar\E^I \E_I$ of the commuting spinors in Lorentzian signature is indefinite, and there is no reason why that norm should be, say, positive everywhere. In other words, there is room for a much larger class of geometries than we will consider here, where we only consider conformally Einstein manifolds (for which the norm of the spinors is everywhere positive). It would be very interesting to examine if $U$ will be a conformal Killing vector that commutes with $V$ for all such Lorentzian geometries. But this goes beyond the scope of our present presentation. However, one can easily see that if $- i \bar\E^I \E_I$ is nonvanishing locally around some point, then we can make a Weyl transformation locally such that we get $- i\bar\E^I \E_I = 1$, from which we can conclude that the manifold has to be conformally Einstein at least locally around such a point. 

\section{Summary of results in ref \cite{Baum}}
In this appendix we translate the results in \cite{Baum} into our notation, which concern chiral conformal Killing spinors on Lorentzian six-manifolds. We also explain how our approach differs from \cite{Baum}. Since \cite{Baum} considers complex spinors, let us start with defining a complex spinor out of our two real spnors $\eps_I$ for $I =1,2$ as $\eps = \eps_1 + i \eps_2$. Proposition 3.1 (2) in \cite{Baum} says that $V_M = 0$ if and only if $\eps_I = 0$. To show this, we note that $\bar\eps \Gamma_0 \eps = i \bar\eps^I \Gamma_0 \eps_I = i V_0$ and further $\bar\eps \Gamma_0 \eps = \eps^{\dag} \eps$ is nonnegative and zero if and only if $\eps = 0$. Proposition 4.2 in \cite{Baum} says that if the Dirac current $\bar\eps \Gamma_M \eps$ is lightlike, then $\Gamma_M \eps V^M = 0$ and $\bar\eps \eps = 0$. This is shown to be true both for chiral and nonchiral $\eps$. In our context with chiral $\eps$, the latter relation can be shown from $P_+ \eps = \eps$ and $\bar\eps P_+ = - \bar\eps$. The result that $\bar\eps \eps = 0$ in \cite{Baum} is the starting point for the classification of Lorentzian geometries, which are not conformally related to Einstein manifolds. We circumvent the result in proposition 4.2 by considering a nonchiral compex conformal Killing spinor $\E$ whose Dirac current $\bar\E \Gamma_M \E$ is not lightlike in which case $\bar\E \E$ is generically nonvanishing. Lemma 5.1 says that for chiral $\eps$ the Dirac current $V_M = \bar\eps \Gamma_M \eps$ is such that $\Gamma_M \eps V^M = 0$ and $V_M$ is lightlike. Theorem 5.1 says the manifold is either Fefferman if $R_{MN} V^M V^N = 0$ or Brinkmann with $\nabla_M \eps = 0$ if $R_{MN} V^M V^N > 0$ where it is also noticed that $R_{MN} V^M V^N$ is constant and nonnegative (by proposition 4.4). The manifolds on which we have a chiral $\eps$ are therefore not conformally Einstein manifolds, (in particular it is impossible to have $R_{MN} V^M V^N > 0$ for an Einstein manifold as $V$ is lightlike.) However, in Euclidean signature they are all conformal Einstein manifolds \cite{baumsbook}.

\end{document}